# Handcrafted vs. Deep Radiomics vs. Fusion vs. Deep Learning: A Comprehensive Review of Machine Learning -Based Cancer Outcome Prediction in PET and SPECT Imaging

Mohammad R. Salmanpour[1,2,3*†], Somayeh Sadat Mehrnia[4†], Sajad Jabarzadeh Ghandilu[5ł], Sonya Falahati[3], Shahram Taeb[6], Ghazal Mousavi[7], Mehdi Maghsoudi[3], Ahmad Shariftabrizi[8], Ilker Hacihaliloglu[2,9], Arman Rahmim[1,2]

[1]Department of Integrative Oncology, BC Cancer Research Institute, Vancouver, BC, Canada
[2]Department of Radiology, University of British Columbia, Vancouver, BC, Canada
[3]Technological Virtual Collaboration (TECVICO Corp.), Vancouver, BC, Canada
[4]Department of Integrative Oncology, Breast Cancer Research Center, Motamed Cancer Institute, ACECR, Tehran, Iran
[5]Department of Electrical Engineering, Sharif University of Technology, Tehran, Iran
[6]Department of Radiology, Guilan University of Medical Sciences, Rasht, Iran
[7]School of Electrical and Computer Engineering, University of Tehran, Tehran, Iran
[8]Department of Radiology, University of Iowa Carver College of Medicine, IA, US
[9]Department of Medicine, University of British Columbia, Vancouver, BC, Canada

(†) Co-First Authors: These authors had equal contributions as co-first authors.

(*) Corresponding Author: Mohammad R. Salmanpour, PhD; msalman@bccrc.ca

## Abstract:

Machine learning (ML), particularly deep learning (DL) and radiomics-based approaches, has emerged as a powerful tool for cancer outcome prediction using PET and SPECT imaging. However, the comparative performance of different techniques—handcrafted/deep radiomics features (HRF/DRF), DL models, and hybrid fusion models (combinations of DRF, HRF, and clinical features)—remains inconsistent across clinical applications. This systematic review analyzed 226 studies published between 2020 and 2025 that applied ML to PET or SPECT imaging for cancer outcome prediction tasks. Each study was evaluated using a 59-item framework addressing dataset construction, feature extraction methods, validation strategies, interpretability, and risk of bias. We extracted key data, including model type, cancer site, imaging modality, and performance metrics such as accuracy and area under the curve (AUC). PET-based models (95%) generally outperformed SPECT, likely due to superior spatial resolution and sensitivity. DRF models achieved the highest mean accuracy (0.862±0.051), while fusion models attained the highest AUC (0.861±0.088). ANOVA revealed significant differences in accuracy (p=0.0006) and AUC (p=0.0027). Despite these promising findings, key limitations remain, including poor management of class imbalance (59%), missing data (29%), and low population diversity (19%). Only 48% adhered to IBSI standards. Standardization and explainable AI are critical for future clinical translation.

## 1. Introduction

Despite significant advances in imaging and artificial intelligence, developing accurate, reliable, and generalizable prognostic models for cancer remains a major clinical challenge. Cancer is the leading cause of death globally, posing immense physical, emotional, and economic burdens on individuals and healthcare systems alike(Siegel, Kratzer et al. 2025).Tumor heterogeneity, recurrence, and variable responses to treatment demand precise diagnostic and predictive tools to guide individualized clinical decisions. In this context, nuclear medicine imaging (NMI)—particularly positron emission tomography (PET) and single-photon emission computed tomography (SPECT)—has emerged as a powerful approach for capturing functional and molecular-level information critical to cancer care.

PET imaging utilizes radiotracers such as 18F-fluorodeoxyglucose (18F-FDG) to generate 3D maps of glucose metabolism, enabling sensitive detection, staging, and treatment monitoring of tumors (Czernin, Allen-Auerbach et al. 2013). In comparison, SPECT imaging, which commonly uses Technetium-99m, provides physiological and functional insights but is generally less sensitive than PET for detecting malignancy (Rahmim and Zaidi 2008, Alqahtani 2023). Both modalities play a central role in modern oncology, particularly when paired with advanced image analysis tools.

Linking imaging data to clinical outcomes began in the 1960s with pattern recognition, evolving into quantitative analysis by the 1980s for CAD applications. Driven by personalized medicine needs, early studies explored ultrasound-CT-PET correlations with malignancy, gene expression, and treatment response (Hatt, Krizsan et al. 2023), radiomics has gained attention as a non-invasive method to extract high-dimensional, quantitative features from medical images (Lambin, Rios-Velazquez et al. 2012). These features—reflecting tumor intensity, shape, texture, and spatial heterogeneity—can reveal clinically relevant phenotypes and support predictions about treatment response and patient survival. Radiomics features can be broadly divided into handcrafted radiomics features (HRFs) and deep radiomics features (DRFs), each offering distinct advantages and challenges.

HRFs are explicitly defined using mathematical formulas and are extracted from specific regions of interest (ROIs). Their interpretability, reproducibility, and alignment with known biological phenomena make them appealing for clinical translation. Applied to PET/CT and SPECT data, HRFs have shown promising utility in predicting tumor metabolism, hypoxia, heterogeneity, and even non-oncologic conditions such as dementia (Gatenby, Grove et al. 2013, Li, Jiang et al. 2019, Chen, Luo et al. 2021, Mu, Jiang et al. 2021, Zhen, Chen et al. 2021, Li, Su et al. 2024).

In contrast, DRFs are derived automatically through deep learning (DL) models such as convolutional neural networks (CNNs) and autoencoders (Nensa, Demircioglu et al. 2019). These data-driven features are capable of capturing abstract, hierarchical representations of tumor characteristics that may be invisible to human observers or traditional feature engineering. DRFs have been successfully used to enhance prognostic performance, particularly in large datasets (Vial, Stirling et al. 2018). However, their clinical adoption is hindered by key challenges—including a lack of interpretability, dependence on large annotated datasets, and reduced effectiveness in rare cancers with limited data.

End-to-end DL models go one step further by combining feature learning and prediction into a single unified framework, eliminating the need for manual feature extraction. These models have achieved state-of-the-art performance in tasks like image segmentation, lesion detection, and outcome prediction. Yet, they are often criticized for their "black-box" nature, high computational requirements, and limited generalizability across institutions or populations (Ahmed, Alam et al. 2023). As a result, hybrid approaches that combine HRFs, DRFs, and clinical or genomic data are increasingly being explored as more robust and interpretable solutions for predictive modeling in oncology.

Despite rapid progress, there is a notable gap in the literature comparing the relative strengths and limitations of HRFs, DRFs, and DL models—particularly in the context of PET and SPECT imaging for outcome prediction across different cancers. Most prior reviews have addressed these methods in isolation, without systematically evaluating their comparative performance or integration potential within nuclear medicine workflows.

Recent literature highlights this growing intersection. For example, Arabi et al. (Arabi, AkhavanAllaf et al. 2021) discussed how DL and radiomics enhance diagnostic accuracy in PET and SPECT by improving image segmentation and quantification. Piñeiro-Fiel et al. (Piñeiro-Fiel, Moscoso et al. 2021) reviewed 290 PET radiomics studies and identified key limitations such as small sample sizes and lack of methodological standardization. Jimenez-Mesa et al. (Jimenez-Mesa, Arco et al. 2023) examined how ML and DL can optimize imaging protocols and identify biomarkers through multimodal data fusion. Similarly, Cheng et al. (Cheng, Chen et al. 2025) and Zhang et al. (Zhang, Liu et al. 2025) emphasized DL's role in image enhancement and lesion detection. Lee et al. (Lee and Lee 2018) underscored the added value of PET radiomics in assessing tumor heterogeneity and guiding treatment strategies.

To address this critical knowledge gap, this review systematically evaluates and compares handcrafted radiomics features, deep radiomics features, and deep learning models for outcome prediction in oncology using PET and SPECT imaging. Specifically, this work investigates: What is the comparative performance of DL models and DRFs versus traditional HRFs in predicting patient outcomes using PET/SPECT imaging data? To what extent can DL algorithms improve the accuracy of treatment response predictions in cancer patients undergoing nuclear medicine therapies? What are the main challenges in standardizing DRFs and DL models across different cancer types in PET/SPECT imaging? How does the integration of multimodal data (e.g., combining PET/SPECT with CT or MRI) enhance the predictive power of ML models in nuclear medicine?

## 2. Materials and Methods

This study conducted a comprehensive and systematic review from 2020 to 2055. The reporting of this review follows the PRISMA-P (Preferred Reporting Items for Systematic Reviews and Meta-Analyses) guidelines (see Figure 1). The protocol for this systematic review was registered with PROSPERO (Registration ID: CRD42024613207) in November 2024. The study selection criteria and the methods used to gather these studies are outlined and will be further explained in the following sections.

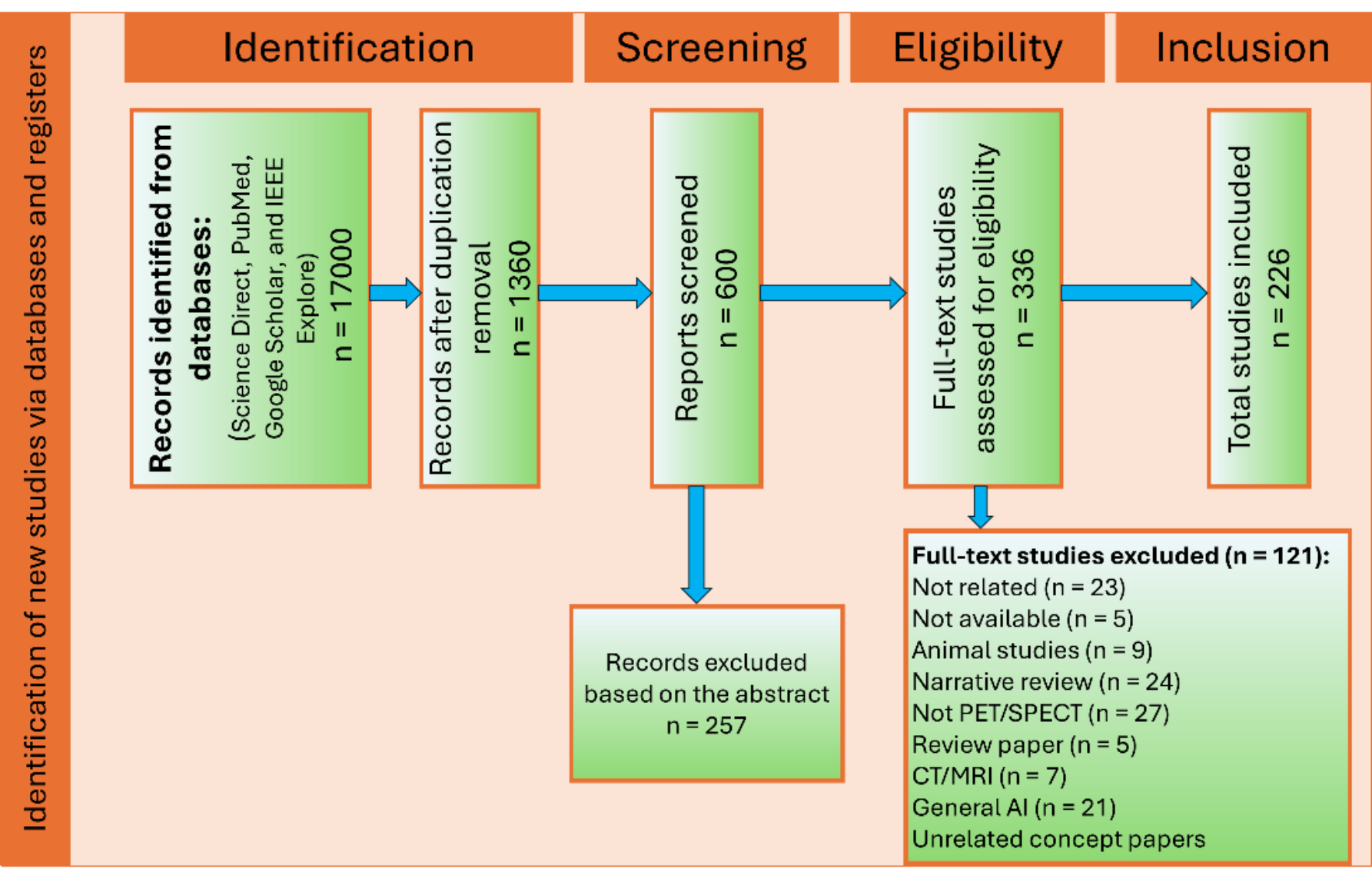

**Fig 1.** PRISMA (Preferred Reporting Items for Systematic Reviews and Meta-Analyses) Flow Diagram: Identification, Screening, Eligibility, and Inclusion of Studies in the Review**.** This diagram illustrates the process of study selection for the review, starting from the identification of records through databases (Science Direct, PubMed, Google Scholar, and IEEE Explore), followed by the removal of duplicates, screening of reports, and full-text assessment. A total of 231 studies were included in the final review after excluding irrelevant studies based on eligibility criteria.

**2.1. Search Strategy:** To systematically identify relevant studies on DL, HRFs, and DRFs applied to cancer diagnosis using PET and SPECT imaging, a comprehensive literature search was conducted for the period between January 1, 2020, and March 30, 2025. Searches were performed across four major academic databases: PubMed, ScienceDirect, Google Scholar, and IEEE Xplore, chosen for their extensive coverage of biomedical and technical publications. Advanced Boolean search strategies were used to refine the results. The core search query included: ("Positron Emission Tomography" OR "Single Photon Emission Computed Tomography" OR "PET" OR "SPECT") AND ("radiomics" OR "texture" OR "textural" OR "deep learning") AND ("cancer" OR "tumor"). To ensure relevance, articles were only considered if their titles included terms related to "Radiomics OR Deep Learning" and "PET OR SPECT." Abstracts were screened for studies focused specifically on prediction or classification tasks in oncology.

**2.2. Eligibility Criteria:** This study applied the following eligibility criteria to ensure the selection of appropriate and relevant research articles for inclusion in the review. (i) Inclusion Criteria: Studies were considered eligible if they: (1) included at least two of the defined search terms; (2) involved human subjects and addressed at least one specific cancer type; and (3) employed PET or SPECT imaging modalities as part of the analysis; (ii) Exclusion Criteria: Studies were excluded if they: (1) did not incorporate PET or SPECT imaging; (2) focused on non-cancer-related conditions; (3) were based on preclinical or animal models; (4) utilized only phantom or simulated data; (5) were non-peer-reviewed formats such as case reports, literature reviews, poster presentations, or conference abstracts; (6) were not written in English; or (7) lacked implementation of AI-based classification or prediction methods.

**2.3. Study Selection and Data Collection:** Eligible studies were systematically recorded in a Microsoft Excel database, which was used to manage the workflow throughout the review process, including initial screening, full-text assessment, and data extraction. For each study, key metadata were documented, including cancer type (e.g., lung, breast, brain, liver), imaging modality (PET or SPECT), and type of modeling approach (HRFs, DRFs, or end-to-end DL models).

**2.4. Data Analysis:** Two independent reviewers assessed the full texts of all included studies. Cancer types accounting for more than 5% of total studies were analyzed separately in the Results section, allowing for in-depth discussion. Less prevalent cancer types were grouped under a collective "Other" category to maintain analytical clarity and focus.

**2.5. Bias Evaluation Methodology:** To rigorously assess the methodological quality and potential biases in the selected studies, we applied an evaluation framework adapted from the Transparent Reporting of a Multivariable Prediction Model for Individual Prognosis or Diagnosis (TRIPOD) guidelines. This customized framework focused on the key areas of model validation, performance metric reporting, dataset diversity, transparency, and reproducibility, tailored specifically to DL and radiomics research in oncology.

**2.6. Dataset Description and Scoring:** A total of 226 studies met the inclusion criteria. Each was evaluated using a 59-item checklist reflecting best practices in ML for medical imaging, as shown in Appendix Table A1. A binary scoring system was applied: a score of 1 was given if the criterion was met, and 0 if not. The criteria spanned preprocessing, model selection, validation methods, hyperparameter tuning, data augmentation, class imbalance handling, feature justification, and external validation. The checklist also distinguished between handcrafted and deep features to enable detailed analysis.

**2.7. Aggregated Results:** Aggregate scores were computed to determine the percentage of studies that adhered to each evaluation criterion. This analysis provided a comprehensive view of current research practices, highlighting methodological strengths and identifying areas requiring improvement. The breakdown of scoring helped illuminate trends, limitations, and gaps in the existing literature on radiomics and DL applications in PET/SPECT-based cancer outcome prediction.

## 3. Results

The study selection process (Figure 1) initially identified 17,000 publications. After removing three duplicates, 16,997 records remained for screening. Titles and abstracts of 600 records were assessed, and 386 were excluded based on the predefined inclusion and exclusion criteria. The full texts of the remaining 365 studies were reviewed, and 134 articles were excluded due to incomplete information or failure to meet eligibility requirements. In total, 226 full-text articles were included in this review. The review focuses on the application of radiomics and deep learning (DL) models for cancer prediction within nuclear medicine imaging. Of the 226 studies, 215 (95.2%) investigated PET, while only 11 (4.8%) focused on SPECT, highlighting PET's dominant role in cancer diagnosis, prognosis, and treatment prediction. PET's ability to generate high-resolution functional images enables the extraction of detailed radiomic features essential for predictive modeling. Regarding image dimensionality, 80.31% of studies used 3D images, 11.2% used 2D images, 8.11% employed both 2D and 3D, and 0.39% used a combination of 3D and 4D images—indicating a strong preference for 3D imaging. Among the PET studies, 20.23% utilized public datasets, while 79.77% relied on private datasets, reflecting a greater dependence on institution-specific data in PET-based research.

### 3.1. Trends of HRF, DRF, and DL in PET/SPECT Cancer Imaging

***PET images.*** PET is widely used in cancer diagnosis, staging, treatment monitoring, and recurrence detection, with 18F-FDG as the primary tracer due to its ability to highlight metabolic activity; other tracers like Ga-68 DOTATATE, 11C-methionine, and 18F-choline are used for specific cancers. Its main applications include tumor detection, prognosis assessment, therapy planning, and drug development (Zhu, Lee et al. 2011, Trotter, Pantel et al. 2023). Figure 2 (up) illustrates the distribution of cancer types in PET studies utilizing HRF/DRF radiomics and DL from 2020 to 2025. Lung cancer (32.3%) was the most studied, followed by head and neck cancer, lymphoma, and ovarian cancer (a). The stacked area graph (b) shows a marked increase in publications, particularly between 2023 and 2024, reflecting the growing adoption of advanced PET imaging techniques for cancer prediction and prognosis. Analysis of imaging modalities revealed a clear dominance of PET/CT, used in 64.16% of studies, underscoring its central role in predictive modeling. PET alone accounted for 30.09%, demonstrating its effectiveness as a standalone modality. In contrast, combinations such as PET+CT and CT+PET/CT were less common (1.33% each), while multi-modal approaches like PET/MRI and PET/CT+MRI were used in fewer than 5% of studies. These findings indicate that, despite emerging interest in multi-modal imaging, PET/CT remains the primary modality in clinical predictive modeling studies.

***SPECT images.*** SPECT imaging, which uses gamma-emitting tracers such as Tc-99m, I-131, In-111, and Ga-67, provides functional information and is commonly used for bone scans, metabolic imaging, and lung cancer screening. It plays a key role in diagnosing cancers like bone metastases (Tc-99m MDP), thyroid (I-131), prostate (Tc-99m Sestamibi), and neuroendocrine tumors (somatostatin receptor imaging), as well as lung, breast, lymphoma, ovarian, gastric, gastrointestinal, brain, bladder, esophageal, and pancreatic cancers (Crișan, Moldovean-Cioroianu et al. 2022, Alqahtani 2023). All 11 SPECT and PET studies published between 2020 and 2025 in this review exclusively employed supervised learning, indicating a clear preference for interpretable, label-dependent models. No unsupervised or semi-supervised approaches were identified. SPECT was the dominant modality, with fewer studies using SPECT/CT or SPECT combined with PET. The use of SPECT/CT reflects a growing trend toward hybrid imaging, offering improved diagnostic accuracy by combining anatomical and functional data (button right). This suggests a future shift toward multimodal imaging to enhance prediction, staging, and treatment planning in cancer care. Lung cancer is the most prevalent, accounting for 50% of the cases, followed by gastrointestinal cancer at 25%, head and neck cancer at 12.5%, and brain tumors at 12.5% (Figure 2, Bottom, c). Trends in cancer types in SPECT Imaging Studies (2020-2024) are shown in Figure 2. Bottom, d. This stacked area chart shows the trend in the use of SPECT imaging for different types of cancer from 2020 to 2024. Lung cancer has remained the most studied, showing a significant increase in recent years, followed by gastrointestinal cancer and brain tumors, which saw growth in 2023. Head and neck cancer studies showed fluctuations throughout the period.

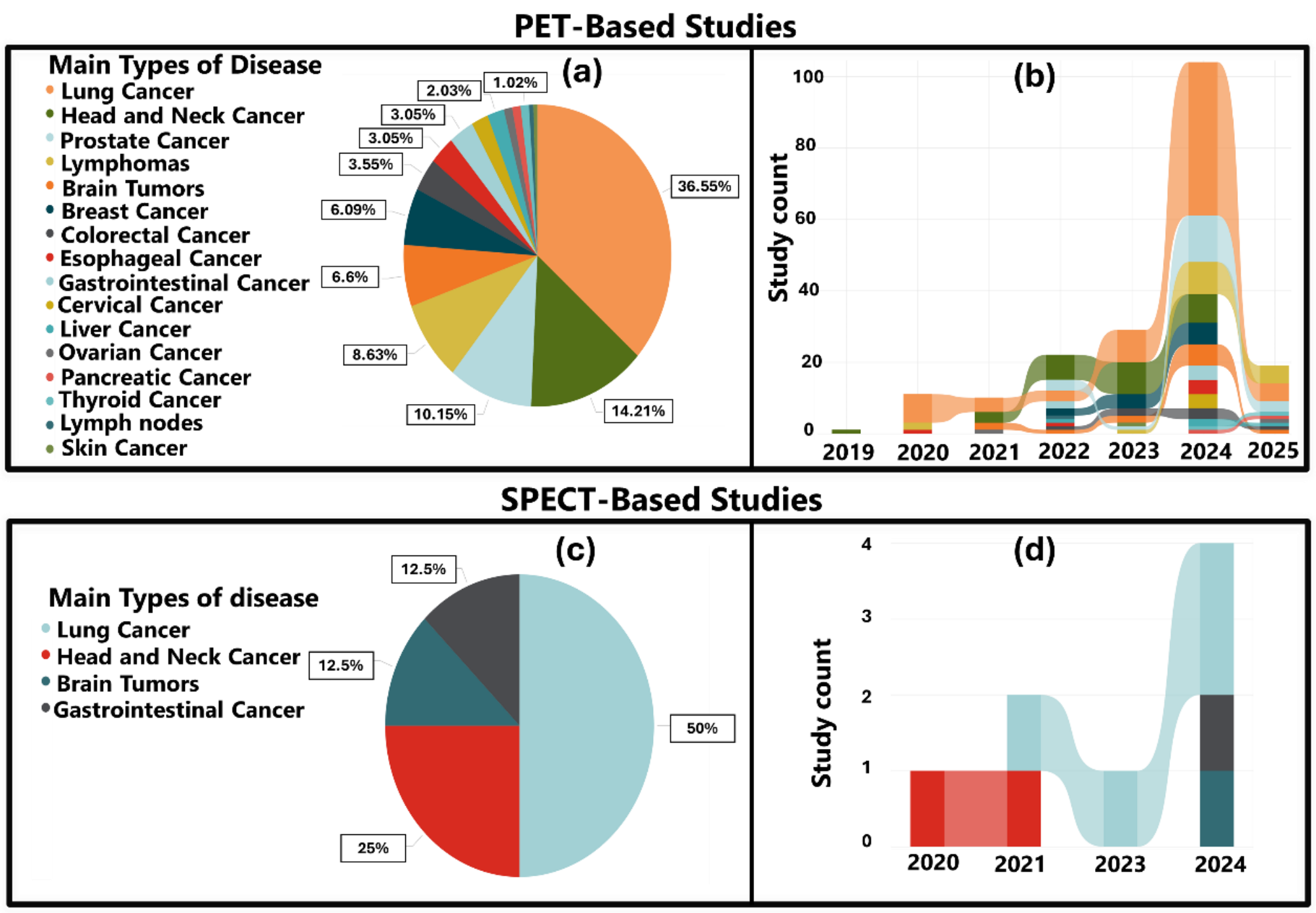

**Fig 2.** Cancer Types in PET (Top) and SPECT (Bottom) Studies Using Radiomics and Deep Learning (2020–2025). The pie chart (left) shows the distribution of cancer types, with lung cancer being the most studied. The stacked area graph (right) illustrates publication trends, highlighting a sharp increase from 2023 to 2024, reflecting the growing adoption of PET/SPECT and AI in cancer prediction.

Figure 3 (left and right) presents an analysis of cancer prediction studies using HRFs, DRFs, and DL models in PET and SPECT imaging from 2020 to 2025, highlighting trends in dataset usage, imaging modalities, and publication venues, respectively. Private datasets dominate, used in 79.11% of studies, while public datasets account for 20.89%. Most studies utilize 3D imaging (79.65%), with limited use of 2D or combined 2D/3D formats. Journal articles are the primary publication venue (79.65%), while conference papers represent 20.35%. Figure 3 (left) shows the distribution of cancer prediction studies using PET-based models: HRFs account for 61%, DL for 23%, and DRFs for 16%. HRFs remain the most widely used approach, though the growing adoption of DL and hybrid methods (e.g., HRF+DRF+DL) reflects a shift toward more comprehensive and automated feature extraction. These trends highlight the potential of combining traditional radiomics with DL to improve prediction accuracy and clinical applicability. Figure 3 (right) illustrates model usage in SPECT-based studies, where HRF still held the largest share (54.6 %), but DL rose to 45.4 %, while DRF was not employed as a stand-alone approach. These patterns indicate a gradual shift from purely handcrafted pipelines toward DL-based strategies, particularly in SPECT, and highlight the persisting under-representation of DRF models across both modalities.

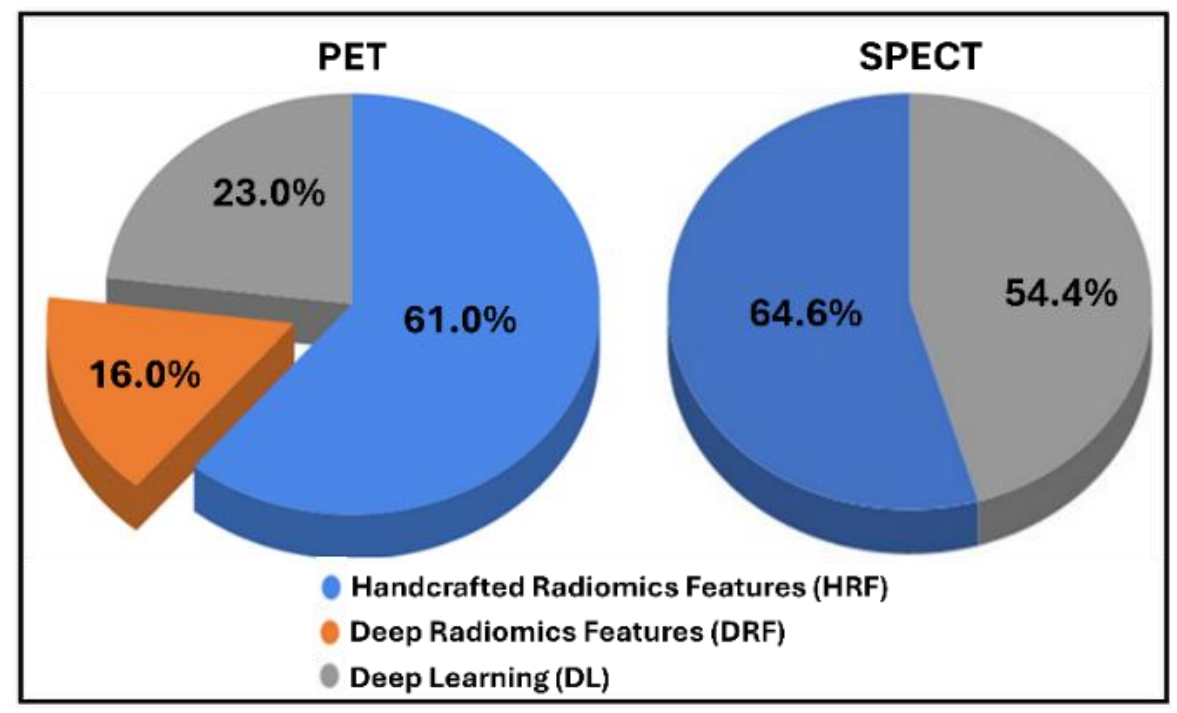

**Fig 3.** Studies on the Use of Handcrafted Radiomics Feature (HRF), Deep Radiomics Feature (DRF), and end-to-end Deep Learning (DL) models in PET/SPECT Imaging for Cancer Prediction (2020-2025). In PET (left), HRF accounts for 61 %, DL for 23 %, and DRF for 16 %. In SPECT (right), HRF represents 54.6 % and DL 45.4 %; no study used DRF alone.

From 2020 to 2025, cancer prediction studies using PET imaging were largely dominated by supervised learning (Figure 4, top a), reflecting a strong preference for interpretable and clinically reliable models over unsupervised or semi-supervised approaches. In ML/AI-based PET image analysis, common evaluation metrics included area under the curve (AUC), accuracy, and sensitivity, with AUC being the most frequently reported (Figure 4, top b). In radiomics-based modeling, LASSO was the dominant algorithm (Figure 4, top c), followed by KNN, Naive Bayes, and Decision Trees, underscoring the continued reliance on established ML techniques in PET-based cancer prediction. Among DL methods, CNNs were most widely used due to their robust feature extraction and predictive performance (Figure 4, top d), while ResNet and attention-based models saw more limited application.

Figure 4 (bottom) provides an overview of SPECT imaging studies during the same period. As shown in Figure 4, bottom b, accuracy and AUC were the most commonly used evaluation metrics, followed by sensitivity and specificity, while F1-score, Negative Predictive Value (NPV), and Positive Predictive Value (PPV) were reported less frequently. Figure 4, bottom c, shows LASSO as the leading algorithm in radiomics applications for feature selection and dimensionality reduction. Other Deep-based methods, such as Attention, CNN, VGGNet, and LightGBM were also used for modeling and optimization. In terms of DL usage (Figure 4, bottom d), CNN was the most dominant architecture, particularly effective in applications such as bone metastasis detection. VGGNet appeared less frequently, indicating lower adoption in SPECT studies.

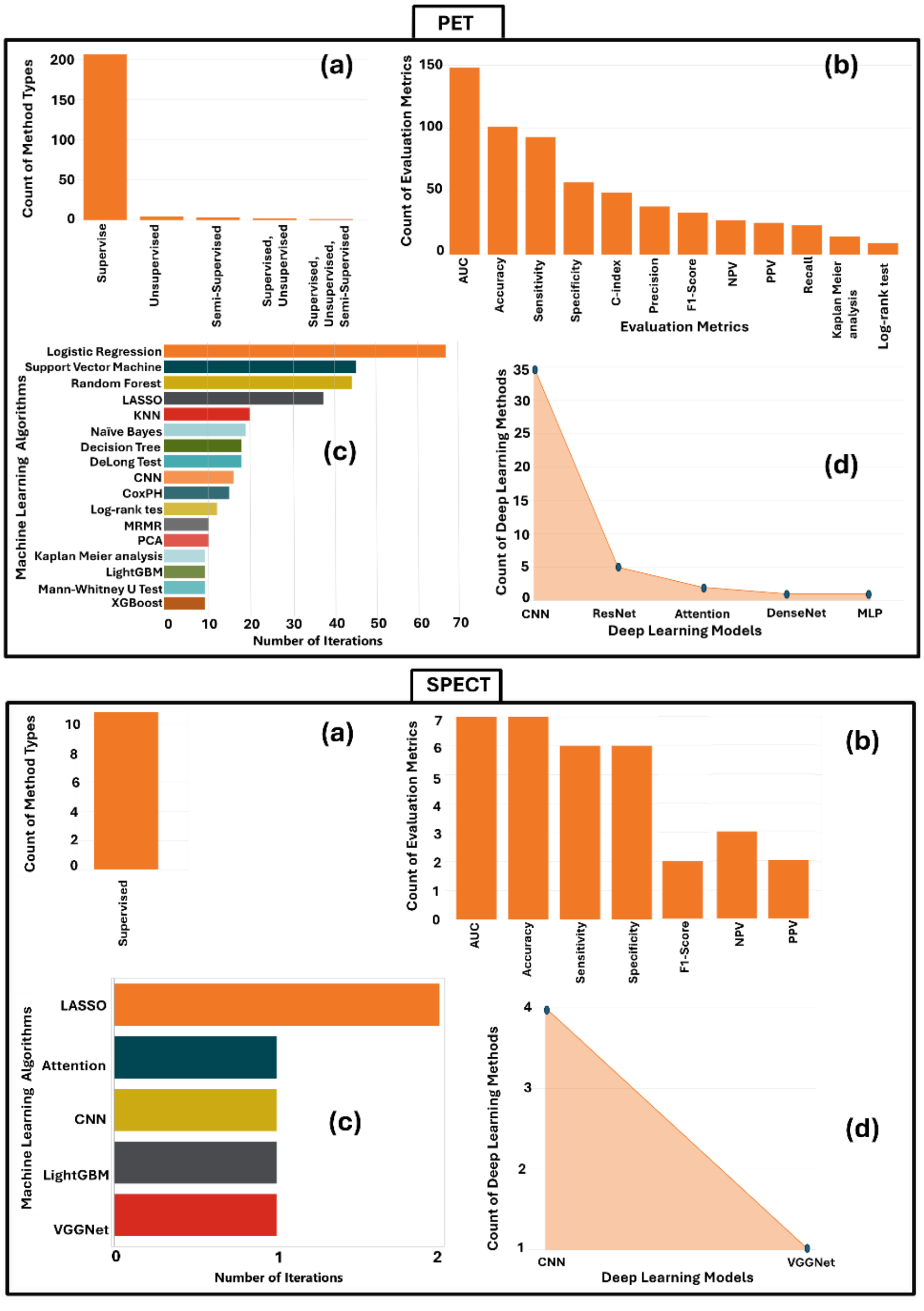

**Fig 4.** Machine learning trends in PET-based cancer prediction (2020–2025).

### 3.1. Analysis of HRF, DRF, and DL in PET/SPECT Cancer Imaging

#### 3.1.1. Comparative Performance Analysis of FRF, DRF, and DL Across Varying Dataset Sizes

As shown in Figure 5, the mean accuracy for DRF models was 0.862 ± 0.051, outperforming all models, including HRF (0.791 ± 0.090), fusion models (a mixture of DRF, HRF, and clinical features; 0.853 ± 0.073), and DL (0.838 ± 0.097). Conversely, fusion models, a mixture of DRFs and HRFs, achieved the highest mean AUC at 0.861 ± 0.088, slightly higher than DRF (0.842 ± 0.082) and DL (0.846 ± 0.097). One-way ANOVA indicated significant differences across model types for both accuracy (F = 6.046, $p$ = 0.0006) and AUC (F = 4.847, $p$ = 0.0027). As shown in Table 1, Tukey HSD post hoc testing confirmed that both fusion, DL, and DRF models significantly outperformed HRF in terms of accuracy ($p \leq 0.05$). However, no pairwise comparisons for AUC reached statistical significance after correction, despite the numerical advantage observed for fusion models. Notably, no significant differences were found among DRF, DL, and fusion models, all of which consistently outperformed HRF. A key limitation is the relatively small number of DRF studies—less than one-third of the HRF sample—raising concerns that the high accuracy observed for DRF may reflect sampling variability.

To assess performance in smaller-scale studies, we conducted a subgroup analysis limited to studies with fewer than 600 participants (n = 46). The performance hierarchy remained consistent, with DRF achieving the highest mean accuracy (0.860 ± 0.052) and fusion models yielding the highest mean AUC (0.862 ± 0.089). One-way ANOVA remained significant for both accuracy ($F$ = 4.56, $p$ = 0.0047) and AUC ($F$ = 3.30, $p$ = 0.0214), indicating that model type continued to influence outcomes in small-to-medium cohorts. Post hoc tests revealed that both fusion and DL models significantly outperformed HRF in accuracy (Fusion vs. HRF: $p$ = 0.001; DL vs. HRF: $p$ = 0.040), while the DRF vs. HRF comparison approached significance ($p$ = 0.05). No significant pairwise AUC differences were observed.

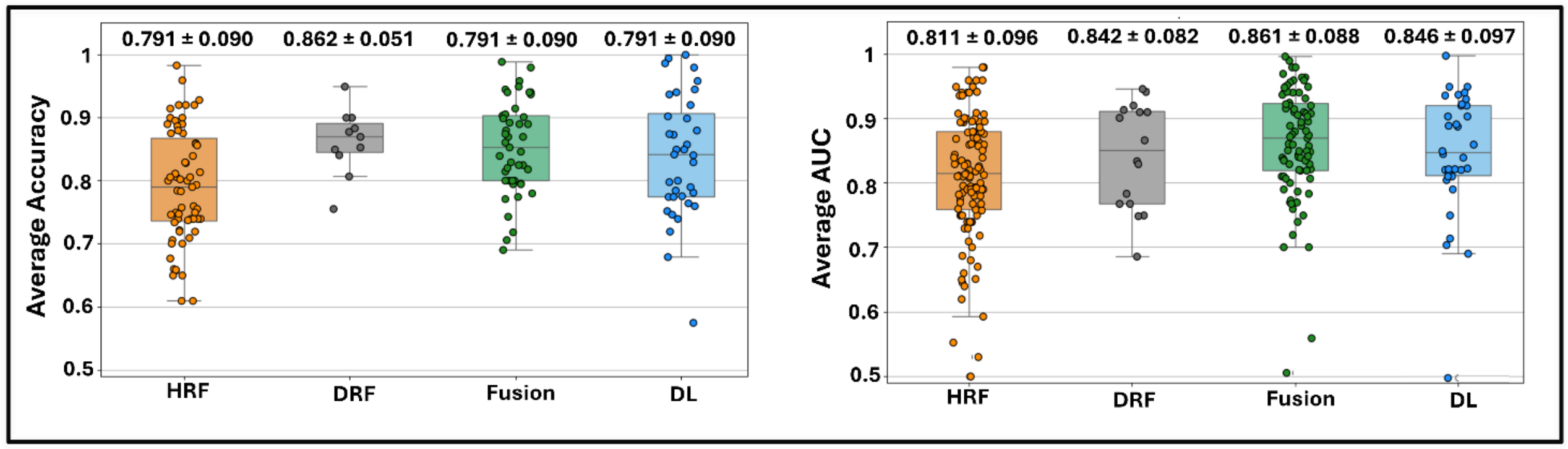

**Fig 5.** Average accuracy and area under the curve (AUC) performance metrics with standard deviations (SD) across Handcrafted radiomics (HRF), deep learning (DL), deep radiomics feature (DRF), and Fusion (a mixture of DRF, HRF, and clinical features).

**Table 1.** One-Way ANOVA Results and Post-Hoc Pairwise T-Tests (Bonferroni-corrected) between different frameworks, such as Handcrafted radiomics (HRF), deep learning (DL), deep radiomics feature (DRF), and Fusion (a mixture of DRF, HRF, and clinical features).

| (b) Tukey HSD post-hoc comparison (α = 0.05) | | | | | | |
|---|---|---|---|---|---|---|
| | Average Accuracy | | | Area Under the Curve (AUC) | | |
| | Difference | P-value | Interpretation | Difference | P-value | Interpretation |
| DRF vs. DL | -0.024 | 0.841 | No significant difference | 0.004 | 0.999 | No significant difference |
| DRF vs. Fusion | -0.010 | 0.986 | No significant difference | 0.018 | 0.891 | No significant difference |
| DRF vs. HRF | -0.072 | 0.051 | Borderline (DRF > HRF) | -0.032 | 0.580 | No significant difference |
| DL vs. Fusion | 0.015 | 0.871 | No significant difference | 0.015 | 0.871 | No significant difference |
| DL vs. HRF | -0.048 | 0.040 | DL significantly outperforms HRF | -0.035 | 0.212 | No significant difference |
| Fusion vs. HRF | -0.062 | 0.001 | Fusion significantly outperforms HRF | -0.018 | 0.433 | No significant difference |

These findings suggest that in smaller datasets, the accuracy advantage of fusion and DL models persists, DRF retains the highest numerical accuracy without clear statistical dominance, and discriminative power (AUC) converges across model types (see Figure 6). Comprehensive accuracy and AUC values, along with dataset sizes for all PET and SPECT studies. Overall, as shown in Supplemental File 1 (Supplemental Tables S1 and S2), DRF-based models were generally trained on smaller sample sizes than other model categories in this study yet achieved performance comparable to DL and fusion-based models. Notably, fusion models—which combine both DRF and HRF—exhibited consistently strong results across varying sample sizes, with particularly robust performance in smaller datasets. These findings suggest that DRF-based approaches, whether used independently or within fusion frameworks, offer superior performance—even in limited sample sizes—when compared to DL-based models.

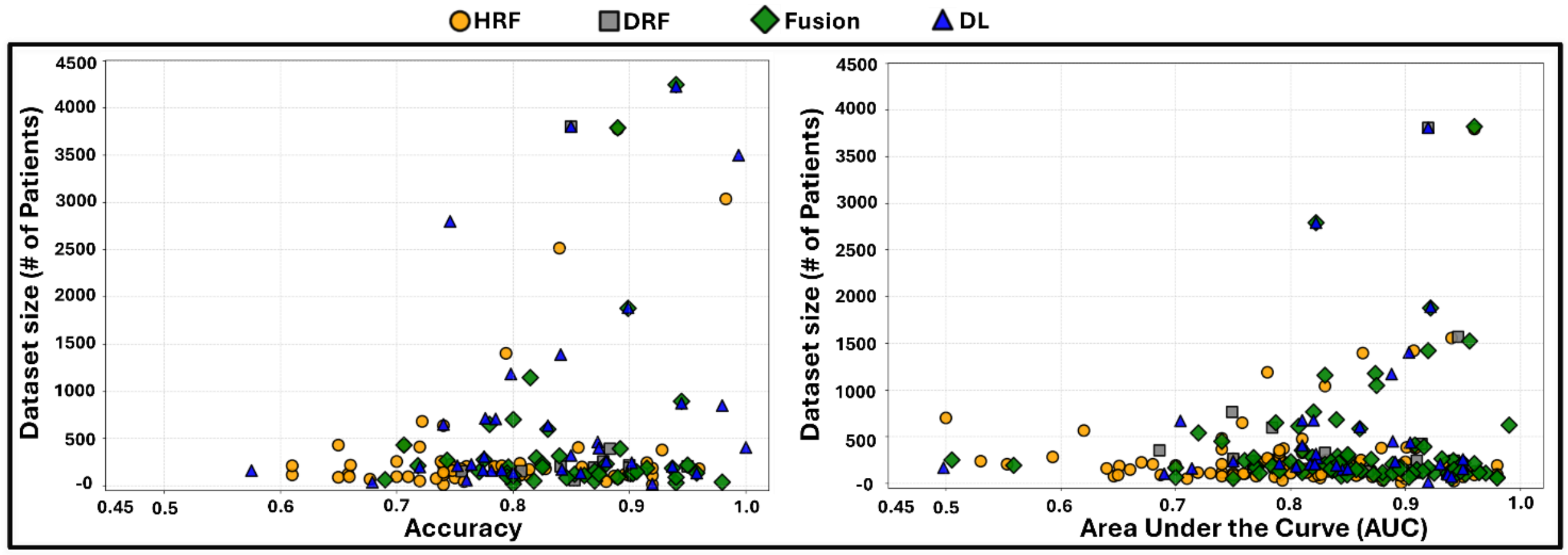

**Fig. 6.** Model Performance vs. Dataset Size: Accuracy and Area Under the Curve (AUC). The left and right sub-figures illustrate the relationship between dataset size and model performance, using Accuracy and AUC, respectively. Each point represents a study using one of the following data modalities: Handcrafted Radiomics Features (HRF), Deep Radiomics Features (DRF), Deep Learning (DL), or Fusion (a mixture of DRF, HRF, and clinical features).

### 3.1.1. PET images

***Lung Cancer.*** From 2020 to 2025, PET imaging has remained central to lung cancer diagnosis, staging, and monitoring due to its superior accuracy over CT, particularly for small lesions (sensitivity: 96%, specificity: 79%, accuracy: 91%) (Vansteenkiste and Stroobants 2006). Studies using HRFs demonstrate their value in quantifying tumor heterogeneity and predicting outcomes in NSCLC. Hosseini et al. (Hosseini, Hajianfar et al. 2021) improved recurrence prediction with sub-volume extension (AUC = 0.65), while Nemoto et al. (Nemoto, Saito et al. 2024) and Wang et al. (Wang, Chen et al. 2024) confirmed HRF-based models' utility for recurrence and EGFR mutation status—though scanner variability affected generalizability. Zhang et al. (Zhang, Liu et al. 2024) showed HRFs could noninvasively differentiate adenocarcinoma from squamous cell carcinoma. These and other works (Amini, Nazari et al. 2020, Kim, Cho et al. 2021, Fujarewicz, Wilk et al. 2022, Tong, Sun et al. 2022, Wilk, Borys et al. 2022, Al-Battat 2024, Andrew William, Ohm et al. 2024, Ciarmiello, Giovannini et al. 2024, Hosseini, Hajianfar et al. 2024, Huang, Cao et al. 2024, Li, Hu et al. 2024, Liu, Sui et al. 2024, Lucia, Louis et al. 2024, Salimi, Hajianfar et al. 2024, Sui, Su et al. 2024, Wang, Bao et al. 2024, Wang, Yang et al. 2024, Yang, Li et al. 2024, Yu, Zhang et al. 2024, Yu, Zhu et al. 2024, Zheng, Hao et al. 2024, Zuo, Liu et al. 2024, Zuo, Liu et al. 2024, Stüber, Heimer et al. 2025, Wang, Dai et al. 2025, Zhai, Li et al. 2025) underscore HRFs' interpretability and clinical relevance. However, DRFs—automatically learned by DL networks—have increasingly outperformed HRFs, particularly when fused with clinical data. Fathi Jouzdani et al. (Fathi Jouzdani, Abootorabi et al. 2024) showed DRF models yielded higher accuracy for overall survival prediction (MAE = 0.38±0.03, c-index = 0.82) than HRFs. Duan et al. (Duan, Zhang et al. 2025) achieved the best AUC (0.853) for lymph node metastasis prediction using combined clinical, HRF, and DRF inputs. Huang et al. (Huang, Zhu et al. 2023) fused lung-brain DRFs using a 3D network to predict brain metastases (AUC = 0.95). Gorgi et al. (Gorji, Hosseinzadeh et al. 2023) demonstrated a 26.5% improvement in survival prediction using semi-supervised learning on PET-HRFs (MAE = 1.55), while Salmanpour et al. (Salmanpour, Gorji et al. 2024) found semi-supervised models with HRFs outperformed both DRFs and supervised models.

Studies directly comparing HRF vs. DRF report similar trends. Gorji et al. (Gorji, Hosseinzadeh et al. 2023) showed DRFs from cropped PET images outperformed HRFs in survival prediction tasks. Huang et al. (Huang, Wang et al. 2022) demonstrated that a hybrid model combining HRF, DRF, and clinical features (AUC = 0.91) outperformed individual HRF (0.82) and DRF (0.90) models. Li et al. (Li, Su et al. 2024) reported the highest AUC (0.954 train, 0.910 validation) for PD-L1 prediction using a fusion model. These results suggest that hybrid modeling offers the best predictive accuracy. In parallel, DL approaches have shown strong performance without relying on explicit feature extraction. Ju et al. (Ju, Li et al. 2024) combined CNN-derived features with whole-body metabolic tumor volume to surpass TNM staging (C-index = 0.771). Munir et al. (Munir, Shah et al. 2025) used autoencoders (GSRA-KL) to generate synthetic DRFs, improving gene mutation prediction. Diao et al. (Diao and Jiang 2024) developed an attention-based RA-DL model combining HRFs and DRFs for improved classification, while Rahmim et al. (Rahmim, Toosi et al. 2023) introduced tensor radiomics (TR), enhancing OS prediction across multiple modalities. More advanced DL architectures have expanded capabilities further. Trabesli et al.

(Trabelsi, Romdhane et al. 2024) applied ResNet-50 for lung tumor segmentation and classification. Baelow et al. (Barlow, Chicklore et al. 2024) used transformer-based models for automated TNM staging extraction. Mu et al. (Mu, Jiang et al. 2020, Mu, Jiang et al. 2021) developed deep scores from PET/CT to predict EGFR and PD-L1 expression, correlating with treatment outcomes. Other DL models—e.g., by Gil et al. (Gil, Choi et al. 2023), Bicakci et al. (Bicakci, Ayyildiz et al. 2020), and Da-Ano et al. (Da-Ano, Tankyevych et al. 2024)—demonstrated the value of PET/CT fusion, whole-body analysis, and peritumoral features in subtype prediction and response modeling.

DL frameworks also excel in specialized tasks. Wu et al. (Wu, Li et al. 2024) classified adenocarcinoma subtypes with AUCs up to 0.93. Li et al. [74] proposed MGTA for distant metastasis prediction (AUC = 0.822). Sultana et al. [75] used an ensemble of MobileNetV2, VGG19, and ResNet50, achieving 98.93% accuracy. Zhao et al. (Zhao, Su et al. 2024) used MobileNetV2 for subtype classification (AUC up to 0.767), while Li et al. (Li, Mao et al. 2024) introduced a two-stage multimodal model (AUC = 0.9227) for pulmonary nodule classification. Final comparative studies further confirmed DL's advantage. Han et al. [80] found that the VGG16 model (AUROC = 0.903, accuracy = 0.841) outperformed HRF models (LDA, SVM) for NSCLC subtype differentiation. Ravikumar et al. (Ravikumar, Kumaran et al. 2023) developed a hybrid ML-DL system (DFM-CNN) using morphological segmentation, improving classification speed and accuracy. In summary, while HRFs remain interpretable and clinically accessible, DRFs and DL models—especially when fused with clinical data—consistently demonstrate superior predictive performance in PET-based lung cancer prediction. Hybrid approaches combining HRFs, DRFs, and DL represent the current frontier, offering improved generalizability and prognostic power across diverse tasks.

Cheng et al. (Cheng, Gao et al. 2024) developed and validated a dual-phase ^99mTc-MIBI SPECT/CT nomogram for diagnosing NSCLC. The model was compared with single-phase SPECT and CT radiomics-based models. The nomogram outperformed clinical models and demonstrated comparable diagnostic performance to radiomics models in both early and delayed imaging phases. Notably, the delayed-phase SPECT/CT provided superior diagnostic accuracy, suggesting its potential as a non-invasive diagnostic tool for NSCLC. However, the study emphasized the need for further validation in larger cohorts. Yubo et al.(Yubo, Qiang et al. 2024) introduced a DL model for the automatic detection of bone metastasis in lung cancer patients using SPECT bone scintigrams. The model employed a custom CNN architecture with two sub-networks for feature extraction and classification. It achieved high performance across metrics: accuracy (0.8038), precision (0.8051), recall (0.8039), specificity (0.8039), F1 score (0.8036), and AUC (0.8489). The exclusion of the urinary bladder region, which shows high 99mTc MDP uptake, further enhanced diagnostic accuracy. This DL model outperformed traditional approaches and demonstrates promise for improving automated bone metastasis detection in clinical practice. These studies highlight the potential of both HRF-based and DL-based models in enhancing SPECT/CT interpretation for lung cancer diagnosis and metastasis assessment.

**Breast Cancer.** From 2020 to 2025, PET/CT-based radiomics has played a significant role in breast cancer prediction, particularly through the use of HRFs. Dholey et al. (Dholey, Santosham et al. 2023) developed an ensemble learning model using HRFs to predict pathological complete response (pCR) after neoadjuvant chemotherapy (NAC), achieving 98% balanced accuracy (PET/CT) and 94.74% F1-score—the first such effort on the QIN-Breast dataset. Similarly, Faraji et al. (Faraji 2024) used PyRadiomics and Random Forest classifiers to predict key biomarkers (ER, PR, HER2, Ki67), achieving high AUCs, while Hou et al. (Hou, Chen et al. 2024) reported strong performance (AUC = 0.95 training; 0.83 testing) using intra- and peritumoral HRFs for NAC efficacy prediction. Other HRF-based models have combined radiomics with clinical data to enhance performance. Ding et al. (Ding, Li et al. 2023) predicted Ki67 expression (AUC = 0.90 train; 0.81 validation), and Aksu et al. (Aksu, Güç et al. 2024) integrated HER2 status to improve pCR prediction (AUC = 0.903). Li et al. (Li, Han et al. 2024) showed that a multi-parametric model combining PET/CT HRFs, clinical, and ultrasound features outperformed standalone approaches for axillary lymph node metastasis prediction (AUC = 0.895). Similarly, Gelardi et al. (Gelardi, Cavinato et al. 2024) and Yang et al. (Yang, Ding et al. 2024) demonstrated that HRFs could predict sentinel lymph node (SLN) metastasis with AUCs around 0.887. Eifer et al. (Eifer, Pinian et al. 2022) effectively distinguished malignant from vaccine-related lymphadenopathy using HRFs (AUC = 0.98). However, Gelardi et al. (Gelardi, Cavinato et al. 2024) noted that while HRFs were less effective for pCR classification in cross-validation, they accurately predicted tumor stage (79%).

In contrast, DRFs have been less frequently applied but show promise. Jia et al. (Jia, Chen et al. 2023) used DenseNet with transfer learning to extract DRFs from PET/CT for breast cancer subtype classification. Online fusion of PET and CT features achieved the best accuracy, demonstrating the value of multi-modal deep feature integration. End-to-end DL models offer a third approach, bypassing manual feature engineering. Inglese et al. (Inglese, Duggento et al. 2022) introduced a DL framework using dynamic PET data (time-activity curves, TACs) rather than static SUV values. This method captured spatiotemporal tracer behavior and outperformed traditional SUV-based analysis in lesion discrimination, providing a non-invasive, time-efficient solution for breast cancer diagnostics. In summary, HRFs remain the most widely used and validated, especially when integrated with clinical data. DRFs, while less common, show potential when paired with feature fusion strategies. DL models offer a fully automated alternative and may provide superior spatial and temporal resolution in complex tasks, although broader validation is still needed.

**Prostate Cancer.** Recent studies have demonstrated the growing role of radiomics and DL in PET imaging for prostate cancer diagnosis, risk stratification, and treatment planning. HRF-based models have shown promising results, though findings suggest clinical features still play a dominant role in some contexts. Molin et al. (Molin, Barry et al. 2025) used [68]Ga-PSMA-11 PET/CT HRFs to predict overall survival in metastatic recurrent prostate cancer. While univariable analysis identified 68 significant HRFs, multivariable analysis revealed that clinical data alone provided better prediction (C-index = 0.722) than HRF (0.681) or combined models (0.704). In contrast, several studies demonstrated stronger HRF performance. Siddu et al. (Siddu, Pawar et al. 2022) integrated HRFs with Gleason grading to classify tumor habitats (accuracy = 90%). Stefano et al. (Stefano, Mantarro et al. 2023) used [18F]-PSMA PET/CT radiomics and SVMs to predict high-grade tumors (AUC = 0.75). Luo et al. (Luo, Wang et al. 2024) achieved AUCs of 0.85–0.96 for predicting seminal vesicle invasion using ^18F-PSMA-1007 PET. Pasini et al. (Pasini, Stefano et al. 2024) and Yang et al. (Yang, Wang et al. 2025) showed AUCs of 0.86 and 0.879, respectively, for risk stratification using automated p ipelines. Bian et al. (Bian, Hong et al. 2024) further validated the prognostic value of periprostatic adipose tissue (PPAT) radiomics (AUCs = 0.85 –0.84 across internal/external validation). Several HRF-based models also outperformed radiologists. Qiao et al. (Qiao, Liu et al. 2024) achieved AUC = 0.844 (training) and 0.804 (testing) using 18F-FDG PET/CT radiomics, with added utility for ADT response and metastatic prediction. Bauckneht et al. (Bauckneht, Pasini et al. 2025) and Maest et al. (Maes, Gesquièr e et al. 2024) developed models distinguishing bone metastases from nonspecific uptake with 84.7% accuracy, aiding less-experienced readers. However, mpMRI-based models still outperformed PSMA-PET/CT for extracapsular extension prediction (Pan, Yao et al. 2024).

DRF and DL models have achieved even higher performance. Zhong et al. (Zhong, Wu et al. 2022) built a DL system for whole-body lesion detection (recall = 100%, F1 = 90.6%) on 68Ga-PSMA-11 PET/CT. Holzschuh et al. (Holzschuh, Mix et al. 2023) used a 3D U-Net for GTV delineation (Dice = 0.71–0.82). Li et al. (Li, Imami et al. 2024) developed a transformer-CNN model (Dice = 0.70, AUC = 0.851, F1 = 0.865), outperforming standard CNNs. Huang et al. (Huang, Yang et al. 2024) created a 3D U-Net for whole-body segmentation with strong performance across centers (internal F1 = 0.824; external F1 = 0.837; $R^2 \geq 0.991$). These models enhance speed and reproducibility in assessing metastatic burden. Comparative studies show that DL and DRF-based methods often outperform HRF alone. Kumar et al. (Kumar, Ramachandran et al. 2024) applied CNNs to HRFs extracted from 68Ga-PSMA PET/CT, achieving 80.7% accuracy in detecting clinically significant cancer. Öğülmüş et al. (Öğülmüş, Almalıoğlu et al. 2025) developed a DL model combining HRF and clinical data, outperforming radiation oncologists in predicting lymph noe involvement. Leung et al. (Huang, Yang et al. 2024) compared DRF and DL models for classifying prostate cancer lesions and patients using [18F] DCFPyL PET/CT, achieving AUROCs of 0.87 (lesion), 0.90 (patient), and 0.92 (cancer-specific), demonstrating high diagnostic confidence and clinical utility. In summary, while HRF models provide interpretable and biologically meaningful insights, DL and DRF models—especially when integrated with clinical features—consistently outperform HRFs in prostate cancer prediction. Hybrid and end-to-end DL approaches now represent the leading direction in PET radiomics for non-invasive, precise disease assessment. Furthermore, Kelk et al. (Kelk, Ruuge et al. 2021) developed a model using voxel-based dosimetry and intra-lesion SPECT-based radiomics to assess tumor response in metastatic prostate cancer treated with 177Lu-DOTAGA radioligand therapy. The results showed a 97.4% decrease in tumor burden after four cycles, along with a significant reduction in PSA, indicating a positive treatment response.

***Lymphoma.*** Lymphomas are heterogeneous malignancies originating from immune system cells. ^18F-FDG PET/CT remains a pivotal imaging modality for staging, response assessment, and treatment planning, especially in Hodgkin's and aggressive non-Hodgkin's lymphoma [120]. Recent studies have explored the integration of radiomics and machine learning to enhance diagnostic accuracy and prognostication. HRFs have been applied across multiple studies. Albano et al. [121] demonstrated that PET/CT-derived shape features could differentiate between DLBCL and MALT in primary gastric lymphoma. Similarly, Zhang et al. [122] distinguished lymphoma from benign lymphadenopathy in patients with fever of unknown origin using selected HRFs and Random Forest models. Triumbari et al. [123] and Ortega et al. [124] showed that while HRFs alone were insufficient for predicting PFS or Deauville scores, integrating them with clinical parameters significantly enhanced predictive accuracy. Wen et al. [125] and Hasanabadi et al. [126] further confirmed that combined HRF-clinical models improved treatment efficacy prediction and subtype classification, respectively. The predictive value of HRFs has also been validated in large-scale prognostic studies. Jing et al. [127] developed a Cox-based model integrating 1,328 HRFs and clinical data to predict PFS and OS in DLBCL with extranodal involvement, outperforming clinical-only models. Zhou et al. [128] showed that both manual and semiautomatic segmentation methods yielded HRFs that improved treatment response prediction.

Yang et al. [122] proposed a Radscore derived from baseline PET radiomics, metabolic, and clinical variables, showing superior performance in elderly DLBCL patients. Yousefirizi et al. [129] highlighted the value of delta HRFs in tracking relapses and progression, while Ligero et al. [130] found a 4-feature PET radiomics signature more effective than SUVmax or TMTV in predicting CAR T-cell therapy response. Despite these advances, Guo et al. [131] found that clinical and consolidation treatment features outperformed HRFs and conventional PET metrics in a PET-adapted protocol (AUC = 0.72 vs. 0.62), underscoring the context-dependent utility of radiomics. However, Cui et al. [132] demonstrated that integrating clinical data with PET-based HRFs yielded the most accurate progression prediction in DLBCL (AUC = 0.898; C-index = 0.853). DRFs, extracted automatically via DL networks, address limitations of manual feature engineering. Jiang et al. [127] combined DRFs from ResNet18 and HRFs to create a radiomic signature for identifying transformed follicular lymphoma. Mitura et al. [133] found both HRFs and DRFs comparably effective in Deauville score prediction.

Chen et al. [134] constructed a DRF signature with clinical factors for DLBCL, achieving high C-indices for PFS and OS. Cui et al [REF] introduced a global-local attention-guided DRF extraction approach, demonstrating high performance in lesion classification across PET/CT datasets. A notable hybrid method combined HRFs and DRFs using attention-based RA-DL networks for subtype classification in limited datasets, achieving AUCs up to 0.95 across liver, lung, and lymphoma cohorts [80]. Although it didn't quantify the individual contributions of HRF and DRF, the study emphasized their complementarity in small-sample settings. Fully automated DL frameworks also show promise. Capobianco et al. [135] automated TMTV estimation using DL, achieving strong agreement with expert contours and robust prognostic stratification (PFS HR = 2.3; OS HR = 2.8). Overall, HRFs remain interpretable and clinically useful, particularly when combined with clinical parameters. DRFs and DL models offer improved automation, scalability, and predictive accuracy. Hybrid models combining HRFs, DRFs, and clinical inputs currently yield the best results across various lymphoma subtypes and treatment scenarios. Their integration into clinical workflows offers a promising path toward personalized therapy planning and response monitoring.

***Colorectal Cancer (CRC).*** 18F-FDG PET/CT is commonly used to detect metastasis and monitor treatment response, especially in advanced CRC (de Geus-Oei, Vriens et al. 2009). Wang et al. (Wang, Zhao et al. 2024) used peritumoral HRFs from 18F-FDG PET/CT to predict lymph node metastasis (LNM) in CRC. Their combined HRF-clinical model outperformed clinical and HRF-only models (AUC: 0.85 training, 0.76 validation). Another study (Wang, Hu et al. 2025) built a PET/CT-based nomogram integrating HRFs and clinical staging (pN, pT) to predict disease-free survival (DFS) in stage II/III colorectal adenocarcinoma. HRFs were selected via univariate Cox, LASSO-Cox, and multivariable Cox regression, achieving AUCs up to 0.86. Xu et al. (Xu, Huang et al. 2024) developed an HRF-based model from [18F] FDG PET/CT to predict LNM by quantifying metabolic and density heterogeneity within lymph nodes. Using LASSO and tenfold CV, the model outperformed conventional imaging in internal and external validation. Chang et al. (Chang, Zhou et al. 2022) introduced a PET/CT DRF signature to predict bevacizumab efficacy in RAS-mutant colorectal liver metastases (CRLM). The model performed well (AUC: 0.982 training, 0.846 internal, 0.768 external), but failed in a chemotherapy-only cohort (AUC: 0.534), indicating its specificity for bevacizumab response. Zhang et al. (Zhang, Zheng et al. 2024) created a radiomics-boosted DL model combining HRFs and DRFs from PET/CT to predict synchronous peritoneal metastasis in CRC. Trained in 220 cases, it achieved high AUCs (0.926, 0.897, 0.885, 0.889) across datasets.

***Head and Neck Cancer (H&NC).*** HRFs have been extensively applied in H&NC for prognosis prediction, tumor characterization, and treatment response assessment. Xu et al. (Xu, Abdallah et al. 2023) demonstrated that manual tumor delineation combined with HRFs predicted PFS with a C-index of 0.719. Pepponi et al.(Pepponi, Berti et al. 2024) and Berti et al. (Berti, Fasciglione et al. 2025) used HRFs from ^68Ga-DOTATOC and ^18F-DOPA PET/CT to predict genetic variants and differentiate paraganglioma subtypes, achieving classification accuracies exceeding 90%. Zhong et al. (Zhong, Frood et al. 2021) identified FDG PET-CT HRFs like MTV and SUVmin as strong predictors of early progression (AUC = 0.94). Lv et al. (Lv, Feng et al. 2021) further highlighted the value of peri-tumoral features, while Bianconi et al. (Bianconi, Salis et al. 2024) and Breylon et al. (Breylon, Jack et al. 2024) improved lymph node and recurrence prediction through PET-based HRF modeling. Salmanpour et al. (Salmanpour, Hosseinzadeh et al. 2023) and Avval et al. (Avval, Amini et al. 2022) demonstrated that PET/CT fusion images and tensor HRFs enhanced survival prediction (C-index = 0.66), with Avval's GTF fusion model performing best. Furukawa et al. (Furukawa, McGowan et al. 2024) combined HRFs with clinical data, achieving a C-index of 0.74. In studies of cervical and oropharyngeal cancers, HRFs also demonstrated strong performance (AUC up to 0.983) (Breylon, Jack et al. 2024, Yang, Zhang et al. 2024).

Several studies compared HRFs and DRFs, often extracted using 3D autoencoders. Salmanpour et al. (Salmanpour, Hosseinzadeh et al. 2022, Salmanpour, Hosseinzadeh et al. 2022, Salmanpour, Rezaeijo et al. 2023) introduced hybrid frameworks combining 215 HRFs and 15,680 DRFs from 17 PET/CT fusion variations. The fused tensor DRF models, especially those coupled with ensemble MLP classifiers, outperformed HRF and CNN models, achieving up to 87% accuracy and external test scores of 85.3%. Lv et al. (Lv, Ashrafinia et al. 2020) confirmed that PET/CT image-level fusion models (WF0.6 and WF0.8) improved survival prediction. Lv et al. (Lv, Zhou et al. 2023) proposed a functional-structural sub-region graph convolutional network (FSGCN), improving generalization across centers by modeling intra-tumoral heterogeneity. They proposed FSGCN, a deep learning model that achieved the highest AUC of 0.781 for PFS prediction when combined with clinical features. In comparison, radiomics models (e.g., PETCT_RS) achieved lower AUCs, with the best at 0.757. DL only approaches have also gained traction. Wang et al. (Wang, Lombardo et al. 2022) used a 3D-ResNet to predict distant metastasis and OS, with PET-only models yielding the best results (HCI = 0.82). Abdallah et al. (Abdallah, Marion et al. 2023) achieved a C-index of 0.86 using harmonized CT and clinical data. Kovacs et al. (Kovacs, Ladefoged et al. 2024) and Arabi et al. (Arabi, Shiri et al. 2020) developed automated PET-based segmentation models, with HighResNet achieving Dice scores of 0.87, matching expert precision. In the HECKTOR 2021 and 2022 challenges, Andrearczyk et al. (Andrearczyk, Fontaine et al. 2021, Andrearczyk, Oreiller et al. 2023) benchmarked DL models for tumor segmentation and PFS prediction; the best model reached a Dice score of 0.78 and C-index of 0.723.

***Oropharyngeal cancer.*** In oropharyngeal cancer, both HRFs and DL models have demonstrated clinical utility for non-invasive prediction tasks, including HPV status and survival outcomes. Haider et al. (Haider, Mahajan et al. 2020) developed radiomics models using PET/CT to predict HPV status in oropharyngeal squamous cell carcinoma. The combined PET-CT model outperformed single-modality approaches, achieving an AUC of 0.78. Although these results are promising, the study concluded that radiomics cannot yet replace tissue-based diagnostics in clinical practice. Ma et al. (Ma, Guo et al. 2024) enhanced their previously proposed TransRP model—an integrated CNN and Vision Transformer (ViT) architecture—for RFS prediction and expanded its utility to additional outcomes: locoregional control (LRC), distant metastasis-free survival (DMFS), and OS. Using a dataset of 400 oropharyngeal squamous cell carcinoma (OPSCC) patients treated with (chemo) radiotherapy, TransRP consistently outperformed CNNs in test C-index values for all outcomes. Moreover, incorporating TransRP predictions into a clinical Cox model further improved OS prediction, demonstrating the benefit of combining DL outputs with clinical and radiomic features for risk stratification. Together, these studies illustrate the evolving role of PET/CT-based AI models in oropharyngeal cancer management, with HRFs aiding in HPV classification and DL methods, particularly hybrid architectures like TransRP, enhancing survival outcome predictions.

Ma et al. (Ma, Li et al. 2023) compared HRF and DL models for recurrence-free survival in oropharyngeal cancer, with DL achieving higher C-index (0.7575), while HRF performed better on test data (0.6683). Meng et al. (Meng, Gu et al. 2022) proposed DeepMTS, a multi-task framework combining segmentation and survival prediction (C-index = 0.681). Hybrid and graph-based approaches have shown superior predictive value. Andrearczyk et al. (Andrearczyk, Fontaine et al. 2021) developed a multi-task DL model integrating segmentation and radiomics-based PFS prediction (C-index = 0.723), outperforming standalone HRF (0.695). Peng et al. (Peng, Peng et al. 2024) introduced a Multi-Level Fusion Graph Neural Network (MLF-GNN), achieving a C-index of 0.788 and AUC of 0.807 for PFS using PET/CT HRFs and clinical data, robust across recurrence and metastasis-free survival predictions in a 642-patient cohort. Fujima et al. (Fujima, Andreu-Arasa et al. 2021) used ResNet-based DL on pretreatment FDG-PET images to predict local treatment outcomes in oropharyngeal squamous cell carcinoma (OPSCC), achieving an AUC of 0.85 and outperforming traditional T-stage models. This highlights the transferable value of DL frameworks for PET-based prognostic modeling, which could be extended to GBC with appropriate datasets. In summary, HRFs offer strong baseline performance in H&NC, especially when combined with clinical features. However, hybrid models that integrate HRFs with DRFs and DL—particularly those using fused images, graph neural networks, or attention mechanisms—consistently outperform standalone models. These advances highlight the growing potential of multi-modal and multi-level frameworks for improving prognostication and guiding personalized therapy in head and neck cancer.

***Thyroid Cancer.*** 18F-FDG PET/CT plays a critical role in thyroid cancer imaging, especially for identifying aggressive, poorly differentiated tumors that are iodine-refractory, as well as for monitoring recurrence. HRFs have recently been explored to enhance diagnostic precision in this context. Fan et al. (Fan, Zhang et al. 2024) developed a PET/CT-based HRFs model to predict lymph node metastasis post-surgery. Using LIFEx, they extracted 164 CT and 164 PET features, selected the most relevant using LASSO, and combined them with clinical variables in a nomogram. The model achieved high performance (AUC = 0.864, C-index = 0.915 in the training group), highlighting its value for postoperative risk stratification. Similarly, Lee et al. (Lee, Lee et al. 2025) constructed an HRFs model to differentiate benign from malignant thyroid nodules using 18F-FDG PET/CT. From 960 candidates, nine features were selected via LASSO to build a radiomics score, yielding AUCs of 0.794 (training) and 0.702 (validation), supporting its potential as a non-invasive diagnostic tool. Zhao et al. (Zhao, Zheng et al. 2023) developed a ResNet-34 to diagnose thyroid diseases using SPECT thyroid scintigraphy images. Trained on 3,194 images from three hospitals, the model achieved 94.4% accuracy in internal validation and 93.1% in external validation. DL outperformed both junior and senior nuclear medicine physicians in both speed and accuracy, highlighting its potential as a clinical decision-support tool. Tsujimoto et al. (Tsujimoto, Teramoto et al. 2021) demonstrated that DL can directly classify benign and malignant thyroid regions from SPECT/CT images without explicit feature extraction. Their study showed that classification accuracy was 73% for SPECT alone, 68% for CT alone, and 74% for whole-body planar images. When using fused SPECT/CT, the overall classification accuracy improved to 80%, with 82% accuracy for malignant and 78% for benign cases. These findings suggest that DL models can significantly enhance the diagnostic performance of SPECT/CT in thyroid cancer by providing accurate, automated classification of lesion types.

***Pancreatic Cancer.*** In pancreatic cancer, PET/CT-based radiomics has shown promise in enhancing diagnostic accuracy, prognosis prediction, and treatment response assessment. HRFs have been particularly effective in identifying clinical biomarkers and improving non-invasive stratification. Wang et al. (Wang, Peng et al. 2024) developed a PET/CT HRF model for preoperative prediction of microsatellite instability (MSI), achieving AUCs of 0.82 (training) and 0.75 (validation). Similarly, Mapelli et al. (Mapelli, Bezzi et al. 2024) applied HRFs from [68] Ga-DOTATOC PET to detect lymph node metastases in non-functioning PanNETs, boosting sensitivity from 24% to 77% and achieving 70% balanced accuracy. HRF-based models have also contributed to survival prediction. Kang et al. (Kang, Ha et al. 2024) developed a PET/CT-derived radiomics risk score (Rad-score) for predicting overall survival in PDAC patients, where the combined HRF-clinical model outperformed the clinical model alone (C-index = 0.740 vs. 0.673). Qi et al. (Qi, Li et al. 2025) integrated HRFs with clinical data, CT imaging, gene mutations, and CA199 using ML models (AdaBoost, XGBoost, LSTM) to effectively predict treatment outcomes in locally advanced pancreatic cancer. To compare HRFs with deep radiomics features (DRFs), Wei et al. (Wei, Jia et al. 2023) developed a fusion model using ^18F-FDG PET/CT to distinguish PDAC from autoimmune pancreatitis. Their hybrid model, integrating both HRFs and DRFs via deep learning, outperformed single-modality approaches (AUC = 96.4%, accuracy = 90.1%), illustrating the value of combining handcrafted and deep features. This demonstrates that while HRFs are effective and interpretable, incorporating DRFs and DL-based fusion improves diagnostic precision and generalizability—especially in differentiating clinically overlapping pancreatic pathologies.

***Esophageal Cancer.*** In esophageal cancer, 18F-FDG PET/CT-based radiomics has been widely investigated to improve response prediction, risk stratification, and outcome forecasting. HRFs have been the primary focus in many studies. Beukinga et al. (Beukinga, Poelmann et al. 2022) evaluated 143 HRFs in 199 patients to predict non-response to neoadjuvant chemoradiotherapy (nCRT), achieving moderate external validation (AUC = 0.67). Eifer et al. (Eifer, Peters-Founshtein et al. 2024) further validated the predictive utility of 17 HRFs in stratifying patients' treatment response via unsupervised clustering ($p < 0.01$), while Shen et al. (Shen, Chou et al. 2024) identified specific HRFs, such as skewness (AUC = 0.716), for predicting circumferential resection margin (CRM) involvement. Several studies have integrated HRFs with clinical and anatomical variables to improve prediction accuracy. Kawahara et al. (Kawahara, Nishioka et al. 2024) developed a survival nomogram in ESCC patients using PET, CT, and HRFs, reporting high C-indices (up to 0.92) in external validation. Similarly, Zhou et al. (Zhou, Zhou et al. 2024) combined PET/CT HRFs with body composition metrics like sarcopenia and VATI, achieving strong prediction for PFS (C-index = 0.810) and OS (C-index = 0.806). Huang et al. (Huang, Li et al. 2024) and Mirshahvalad et al. (Mirshahvalad, Ortega et al. 2024) also showed improved OS prediction using nomograms and hybrid models integrating PET/CT HRFs, clinical features, and sarcopenia, with AUCs up to 0.88. In a direct HRF vs. DRF comparison, Yuan et al. (Yuan, Huang et al. 2024) developed a hybrid model incorporating both HRFs and DRFs from PET/CT to predict cervical lymph node metastasis in ESCC. The DRF-clinical model outperformed all other configurations, demonstrating the superiority of DRFs when paired with clinical data for non-invasive risk assessment and treatment decision-making. Collectively, these findings highlight the evolving role of HRFs in prognosis and treatment guidance, while also emphasizing that the integration of DRFs and DL-enhanced models offers superior predictive accuracy—particularly for metastasis detection and personalized treatment planning in esophageal cancer.

***Cervical Cancer.*** In cervical cancer, 18F-FDG PET/CT-based radiomics has demonstrated significant promise for subtype classification, treatment response prediction, and survival assessment. HRFs have been widely explored for these purposes. Liu et al. (Liu, Cui et al. 2024) developed an HRF model using Light GBM to differentiate between squamous cell carcinoma and adenocarcinoma, achieving high accuracy (0.915) and AUC (0.851), outperforming CT-based and combined models. Cepero et al. (Cepero, Yang et al. 2024, Cepero 2024) showed that FDG-PET HRFs could predict early response to chemoradiation, with the midway-treatment model achieving superior performance (AUC = 0.942 vs. 0.853 pre-treatment), highlighting the value of longitudinal radiomics for real-time adaptation. Collarino et al. (Collarino, Feudo et al. 2024), however, found that pretreatment HRFs lacked sufficient predictive power (AUC $< 0.70$) for response and survival in locally advanced cervical cancer (LACC), emphasizing the need for methodological standardization and prospective validation. In contrast, Li et al. (Li, Jin et al. 2021) demonstrated that PET-CT HRFs—including peritumoral features—effectively predicted E-cadherin expression (AUC = 0.915) and pelvic lymph node metastasis, offering a non-invasive approach for prognostic evaluation. Liu et al. (Liu, Cui et al. 2024) integrated HRFs with clinical variables to build a nomogram for predicting PFS in LACC patients, which outperformed models based on clinical or imaging features alone, showing robust 3- and 5-year survival predictions. In a direct comparison of DRFs and traditional approaches, Yang et al. (Yang, Zheng et al. 2023) used a ResNet50-based CNN to classify metastatic vs. lymphomatous cervical lymph nodes, achieving an AUC of 0.845. When SVM models were trained using DRFs, performance improved further (AUC = 0.901, accuracy = 87.0%), indicating that DRFs provided more informative features than DL alone. This underscores the diagnostic advantage of combining DL with DRFs for improving precision in lymph node classification and personalized treatment planning in cervical cancer.

***Ovarian Cancer (OC).*** In OC, [18F]-FDG PET/CT-derived HRFs have shown promising applications in tumor characterization, prognosis, and staging. Wang et al. (Wang, Xu et al. 2022) developed an HRF model using the Habitat method to predict Ki-67 status in high-grade serous ovarian cancer (HGSOC), demonstrating superior predictive performance compared to whole-tumor texture features ($p < 0.001$). This method shows potential for guiding individualized diagnosis and treatment. In another study, Wang et al. (Wang and Lu 2021) constructed a PET/CT-based nomogram integrating HRFs and clinical features to predict PFS in advanced HGSOC. The combined model achieved a C-index of 0.70 in both training and validation cohorts, offering a robust tool for survival prediction. Xu et al. (Xu, Zhu et al. 2025) further demonstrated that combining PET/CT metabolic parameters and HRFs with clinical data using an adaptive ensemble machine learning model could accurately predict FIGO staging in OC (AUC = 0.819), with strong performance in subtypes like OCCC and MCOC (AUC = 0.808). These findings collectively highlight the value of PET/CT HRFs, especially when fused with clinical information, in enhancing OC staging and prognosis modeling.

***Brain Tumors.*** HRFs have been extensively applied in pediatric neuroblastoma and gliomas for diagnostic, prognostic, and molecular characterization. Feng et al. (Feng, Yang et al. 2022, Feng, Yao et al. 2024, Feng, Zhou et al. 2024, Feng, Yao et al. 2025) and developed a series of PET/CT-based HRFs nomograms to predict relapse timing, bone marrow involvement, tumor subtype (e.g., ganglioneuroblastoma vs. neuroblastoma), and INPC classification. These models, integrating clinical risk factors, achieved consistently high performance (AUCs: 0.795–0.932) and demonstrated strong generalizability. Qian et al. (Qian, Feng et al. 2023) confirmed that combining HRFs with clinical data improves INPC-type prediction.Noorman et al. (Noortman, Vriens et al. 2022) used HRFs, SUVmax, and biochemistry to classify genetic clusters in pheochromocytomas/paragangliomas, with the full model reaching AUC = 0.88. In gliomas, Gutsche et al. (Gutsche, Scheins et al. 2021) identified 297 reproducible FET PET HRFs unaffected by IDH genotype. Zhou et al. (Zhou, Wen et al. 2024) built a ^11C-MET PET/CT HRFs model to predict IDH mutation and WHO grade (AUCs: 0.880–0.866). Ahrari et al. (Ahrari, Zaragori et al. 2024) demonstrated that delta-HRFs from dynamic 18F-FDOPA PET predict progression-free survival better than static features (C-index = 0.783). Lohmann et al. (Lohmann, Elahmadawy et al. 2020) used dynamic FET PET HRFs to distinguish pseudoprogression from early progression in glioblastoma with perfect test set sensitivity (100%), outperforming TBRmax. Kaiser et al. (Kaiser, Quach et al. 2024) combined TSPO PET, dynamic FET PET, MRI kurtosis, and age, achieving AUC = 0.97 for IDH prediction.

Comparative studies show the added value of DRFs and DL. Shazadi et al. (Shahzadi, Seidlitz et al. 2024) compared MET-PET-based 3D-DenseNet DRFs (AUC = 0.95) and T1-MRI HRFs (AUC = 0.78), with DRFs showing superior tumor detection and outcome prediction. Vedaei et al. (Vedaei, Mashhadi et al. 2024) fused resting-state fMRI and PET in a DL model for mild TBI diagnosis (AUC = 93.75%). Hossain et al. (Hossain, Qureshi et al. 2023) used dynamic PET and MRI to distinguish recurrence vs. necrosis in glioblastoma (accuracy = 0.74). Usha et al. (Usha, Kannan et al. 2024) introduced TumNet, a fusion CNN achieving 98% accuracy and 96% sensitivity using CT and MRI. Kawauchi et al. (Kawauchi, Furuya et al. 2020) applied CNNs to whole-body FDG PET scans, achieving 99.4% accuracy in benign/malignant classification. Nobashi et al. (Nobashi, Zacharias et al. 2020) optimized window settings for brain PET CNNs, reaching 82% accuracy. Dwarakadeesh et al. (Dwarakadeesh, Shinde et al. 2024) fused CT/MRI using CNNs and VGG19 to improve segmentation and diagnosis. Kundu et al. (Kundu, Terrell et al. 2024) compared HRFs and DL using FDG PET and MRI for progression vs. necrosis classification in GBM; DL slightly outperformed HRFs (accuracy = 0.76 vs. 0.74). Together, these studies highlight that HRFs remain valuable for structured, interpretable analysis, while DRFs and DL models—especially when fused or dynamically optimized—provide higher sensitivity and accuracy in complex neuro-oncology tasks. Integrating all three approaches enhances diagnostic reliability, staging, and personalized treatment planning across pediatric and adult neuro-oncology. Moreover, Feng et al. (Feng, Yang et al. 2024) developed a radiomics nomogram combining 123I-MIBG SPECT-CT and clinical factors to predict event-free survival (EFS) in high-risk pediatric neuroblastoma. The model outperformed clinical models with C-indices of 0.819 and 0.712 in training and validation cohorts.

***Nasopharyngeal Carcinoma (NPC).*** In NPC, Gu et al. (Gu, Meng et al. 2022) developed a 3D CNN-based DRFs model using pretreatment PET/CT to predict 5-year PFS in 257 patients with advanced disease. The DRFs model achieved superior performance (AUC: 0.842 internal, 0.823 external) compared to traditional HRFs (AUC: 0.796 internal, 0.782 external) and single-modality models. The integration of TNM staging with the DRF model further improved prognostic accuracy, and the DRF signature enabled effective risk stratification ($p < 0.001$), unlike clinical models alone ($p = 0.177$ external). These findings highlight the added value of DRFs over HRFs in NPC outcome prediction.

***Hepatocellular carcinoma (HCC).*** In HCC, HRFs have shown promise in outcome prediction and tumor characterization. Fan et al. (Fan, Long et al. 2025) developed 68Ga-FAPI PET-based radiomics models to predict microvascular invasion (MVI), finding that a logistic regression model using HRFs at a 50% SUV max threshold yielded the best performance (AUC = 0.896, accuracy = 87.5%), emphasizing the importance of segmentation strategy. Sui et al. (Sui, Su et al. 2024) used habitat-based radiomics from 18F-FDG PET/CT to build a prognostic model, extracting 4032 features and applying a stacking ensemble learning approach. Their MLP-Cox model, enhanced with clinical data, demonstrated strong predictive capability for survival. For intrahepatic cholangiocarcinoma (IHCC), Kwon et al. (Kwon, Kim et al. 2024) applied unsupervised clustering to ^18F-FDG PET/CT HRFs, identifying radiomics-based subgroups that were significantly associated with recurrence-free and overall survival ($p < 0.0001$). These clusters served as independent prognostic indicators and were linked to distinct molecular pathways, providing complementary insights to genomic profiling. Although direct DRF-based models for HCC were not identified in this subset, Fujima et al. (Fan, Long et al. 2025) demonstrated the effectiveness of DL in oral cavity squamous cell carcinoma using [18F]-FDG PET/CT. Their DL model achieved a diagnostic accuracy of 0.8 and significantly stratified disease-free survival outcomes, supporting the broader utility of DL in PET imaging for solid tumors. These findings highlight the growing utility of HRFs in liver cancers like HCC and IHCC, while also pointing to the emerging potential of DL models for survival prediction, especially when integrated with clinical and molecular data.

**Gastric cancer (GC).** In GC, HRFs extracted from 18F-FDG PET/CT have demonstrated significant potential for non-invasive prediction of clinical and molecular outcomes. Xue et al. *(Xue, Yu et al. 2022)* developed a radiomics-based nomogram combining HRFs and clinical factors to predict lymph node metastasis (LNM). Their multivariate logistic regression model achieved strong predictive performance across training and validation cohorts, highlighting its clinical utility. Chen et al. *(Chen, Zhou et al. 2024)* focused on using HRFs from visceral adipose tissue (VAT) to predict molecular markers like Her-2 and Ki-67 expression. Using PyRadiomics and logistic regression models, their approach achieved high accuracy (Her-2: AUC = 0.84, accuracy = 0.86; Ki-67: AUC = 0.86, accuracy = 0.79), supporting VAT radiomics as a valuable imaging biomarker. Zhi et al. *(Zhi, Xiang et al. 2024)* further integrated HRFs with TNM staging to predict overall survival using a random survival forest model, achieving a C-index of 0.817 (training) and 0.707 (validation), outperforming clinical features alone. In contrast, Huang et al. *(Kun, Gao et al. 2024)* applied DL to improve risk stratification in gastrointestinal stromal tumors (GISTs), generating pseudo-PET images with a conditional PET GAN (CPGAN) and training a transformer-based model (TMGRS). The model achieved high predictive accuracy (0.937), showing that DL can enhance image realism and downstream prediction tasks when PET imaging is limited. Collectively, these studies demonstrate the growing utility of HRFs in prognosis and biomarker prediction for GC and related tumors, while DL techniques—particularly GANs and transformers—offer complementary advances by enriching data quality and model performance for risk stratification.

**Oral Squamous Cell Carcinoma (OSCC).** In OSCC, HRFs extracted from preoperative 18F-FDG PET/CT have shown strong prognostic and predictive utility. Song et al. *(Song, Tian et al. 2024)* developed a PET/CT-based radiomics score (RADS score), identifying it as an independent prognostic factor for OS and progression-free survival (PFS), while sarcopenia showed no significant impact. The integrated model combining RADS score with clinical features achieved robust performance in both training and validation cohorts, supporting its role in individualized risk stratification. Complementarily, Nikkuni et al. *(Nikkuni, Nishiyama et al. 2024)* assessed HRFs for predicting histological grade, extracting 2993 features from PET scans, and applying five machine learning models. The random forest classifier performed best, achieving an AUC of 0.84. These results underscore the promise of HRFs for outcome prediction and histological classification in OSCC, paving the way for personalized treatment planning.

***Renal Cell Carcinoma (RCC).*** Kumar et al. *(Kumar, Shamim et al. 2024)* analyzed 18F-FDG PET/CT images using Haralick texture features and found that mean/median texture values and tumor-to-liver ratios yielded the highest diagnostic accuracy for distinguishing tumoral from healthy renal tissue. Klontzas et al. *(Klontzas, Koltsakis et al. 2023)* developed a machine learning model combining ^99mTc-Sestamibi SPECT/CT with HRFs to differentiate benign from malignant renal oncocytic tumors. The model achieved high diagnostic performance (accuracy = 95%, AUC = 0.983), outperforming traditional imaging approaches. Despite its promising results, the study's small sample size and lack of external validation limit its generalizability. Future research using larger, multi-center datasets is necessary to validate its clinical utility and expand its application to broader renal tumor subtypes.

***Gallbladder Cancer (GBC).*** In GBC, radiomics and DL approaches have shown promise for improving diagnostic precision. Han et al. *(Han, Wang et al. 2024)* developed a radiomics model using 18F-FDG PET/CT to differentiate non-metastatic GBC from cholecystitis. By applying mRMR and LASSO for feature selection, their radiomics model achieved AUCs of 0.940 and 0.906 in the training and testing cohorts, respectively. The combined clinical-radiomics model further improved diagnostic accuracy (AUC = 0.964), demonstrating strong potential for non-invasive differential diagnosis. While direct DL applications in GBC remain limited, related studies suggest high promise.

***Liver Metastases***. In liver metastases, particularly neuroendocrine liver metastases (NELMs) and intrahepatic cholangiocarcinoma (IHC), both HRFs and DL methods have shown diagnostic and prognostic value. Ingenerf et al. *(Ingenerf, Grawe et al. 2024)* identified several imaging and clinical prognostic factors in NELM patients treated with trans-arterial radioembolization (TARE). Lower Ki-67 index, lower hepatic tumor burden, absence of extrahepatic disease, and higher Tmean/Lmax ratios were associated with longer OS, while higher SUVmax predicted better PFS. These HRFs and PET-derived metrics may aid in patient stratification and treatment planning. Fiz et al. *(Fiz, Masci et al. 2022)* demonstrated that HRFs from preoperative 18F-FDG PET/CT can improve the prediction of tumor grading, microvascular invasion, and survival outcomes in IHC. Their combined clinical-radiomic models outperformed models using clinical data alone, supporting the added value of radiomics in preoperative assessment. On the DL front, Xing et al. *(Xing, Silosky et al. 2024)* introduced a single-stage deep learning framework for lesion detection in ^68Ga-DOTATATE PET images of neuroendocrine tumors. The model achieved an F1 score of 83.24%, surpassing state-of-the-art methods and offering a robust, automated tool for lesion identification. Together, these studies underscore the complementary roles of HRFs and DL in improving prognostication and detection in liver metastases, especially when integrated with clinical parameters. Behmanesh et al. *(Behmanesh, Abdi-Saray et al. 2024)* assessed treatment response to 177Lu-DOTATATE in patients with neuroendocrine tumors (NETs) using radiomics features extracted from SPECT and SPECT-CT images, combined with clinical variables. Among various ML models evaluated, the random forest classifier achieved the highest predictive accuracy—up to 83%—when using SPECT-CT data, while models based on SPECT alone performed poorly. The study highlights the added value of SPECT-CT radiomics in conjunction with clinical features for noninvasive and personalized prediction of therapeutic outcomes in NET patients.

***Melanoma.*** Vgenas et al.(Vagenas, Economopoulos et al. 2023) proposed a whole-body segmentation framework for identifying metastatic melanoma (MM) lesions in 3D FDG-PET/CT images. Their decision support system utilizes Ensemble Unsupervised Segmentation, and a region classification model based HRFs and neural networks, achieving high performance with sensitivity of 83.68%, specificity of 91.82%, F1-score of 75.42%, AUC of 94.16%, and balanced accuracy of 87.75%, as confirmed by public dataset evaluation.

***Chondrosarcoma.*** Yoon et al. (Yoon, Choi et al. 2023) used SPECT/CT radiomics to differentiate enchondromas from grade I chondrosarcomas in long bones. The model demonstrated strong diagnostic performance, with sensitivity and specificity ranging from 83.3% to 90.9%. However, the small sample size (n=49), retrospective design, and lack of external validation limit the generalizability of the findings. Jin et al. (Jin, Zhang et al. 2021) developed SPECT/CT radiomics models to distinguish vertebral bone metastases from benign lesions, achieving high AUCs in both training (0.894–0.951) and validation (0.844–0.926) datasets. The models outperformed human experts and offered a promising non-invasive diagnostic alternative, though limitations include the single-center design and lack of histopathological confirmation. Wang et al. (Wang, Qu et al. 2024) created predictive models for bone metastasis in newly diagnosed prostate adenocarcinoma using SPECT radiomics combined with clinical data. Across 176 patients, the models achieved AUCs between 0.87 and 0.98, outperforming clinical assessments and supporting their use for individualized metastasis risk prediction. Lin et al. (Lin, Li et al. 2021) developed deep learning classifiers using thoracic SPECT images, with the VGG-based SPECS V21 model achieving an AUC of 0.993. Similarly, Zhao et al. (Zhao, Chen et al. 2021) applied VGGNets to grayscale SPECT images, attaining high accuracy (0.98) and AUC (0.993). These models surpassed traditional methods in efficiency and performance, though dependence on grayscale images and dataset quality may limit broader application. Collectively, these studies (Jin, Zhang et al. 2021, Lin, Li et al. 2021, Zhao, Chen et al. 2021, Klontzas, Koltsakis et al. 2023, Yoon, Choi et al. 2023, Wang, Qu et al. 2024) demonstrate the promise of SPECT/CT radiomics and DL in bone tumor and metastasis assessment, while highlighting the need for standardized protocols and external validation.

***Multiple types of cancer datasets.*** Hinzpeter et al. (Hinzpeter, Mirshahvalad et al. 2024) demonstrated that HRFs extracted from [18F]FDG PET/CT scans could be effectively used to classify tumors across multiple cancer types—specifically pulmonary, gastroesophageal, and head and neck cancers. By applying recursive partitioning and all-subset regression, their study showed that HRFs could accurately differentiate both histological subtypes and anatomical origins, with particularly strong performance in classifying pulmonary malignancies. These findings support the feasibility of using PET/CT-based HRFs for multi-cancer classification and personalized treatment strategies. Leung et al. (Leung, Rowe et al. 2024) developed a deep semi-supervised transfer learning approach utilizing a U-Net architecture for fully automated whole-body tumor segmentation and prognostication across six cancer types using 18F-FDG and PSMA PET/CT scans. Despite limited annotated data, their model achieved high segmentation accuracy (median Dice scores up to 0.83) and strong prognostic performance (AUC up to 0.86), enabling survival prediction and therapy response assessment. This approach highlights the potential of scalable DL methods for comprehensive cancer evaluation across heterogeneous datasets, particularly when labeled data is scarce.

### 3.2. Bias Evaluation Overview

This analysis evaluates the methodological rigor and bias in ML-based medical imaging studies using 59 criteria. These questions assess key practices in dataset handling, model development, evaluation metrics, reproducibility, and fairness.

Appendix Table A1. summarizes the proportion of studies fulfilling each criterion, with detailed raw data in Supplemental File 2.

***Dataset Handling and Validation.*** A majority of studies (96%) split datasets into training, validation, and test sets, reducing overfitting by ensuring independent evaluation (Q1). However, only 59% addressed class imbalance (Q2), and 67% employed data augmentation for improved generalization (Q3). Systematic model comparison was conducted in 95% of studies (Q4), while 84% used proper cross-validation, including techniques like leave-one-center-out (Q5). Hyperparameter tuning (Q6) and model stability evaluation (Q7) were each performed in 76 and 63% of studies. Multicenter data were used for training in 67% (Q8) and for external testing in 71% of studies (Q9), enhancing generalizability.

***Evaluation Metrics and Reporting.*** Metric reporting was extensive: 82% of studies reported ≥6 metrics (Q10), 84% reported ≥5 (Q11), and 88% reported ≥4 (Q12). Furthermore, 92% reported ≥3 metrics (Q13), 97% reported ≥2 (Q14), and all studies reported at least one (Q15). Outcome definitions were consistent in 91% of studies (Q16), and performance comparisons were done using the same test set in 97% (Q17). Transparent reporting of model performance—including both strengths and limitations—was noted in 88% (Q18)

***Dataset Diversity and Generalizability***. Dataset diversity was addressed in 82% of studies (Q19), but only 9% considered socioeconomic/geographic variation (Q20), and 19% included race/population variability in comparisons (Q25). Eligibility criteria and treatment details were provided in 65% (Q21), while missing data and its impact were discussed in just 29% (Q22). Differences between training and evaluation sets were explicitly identified in 75% (Q23). Harmonization of multicenter data or scanners was performed in 57% (Q24), and 81% discussed generalizability to unseen data (Q26). Finally, 75% of studies included sufficient detail for replication (Q27).

***Imaging and Ground Truth.*** Ground truth labeling by multiple experts was employed in 55% of studies to reduce subjectivity (Q28). Diverse scanner types and protocols were used in 69% (Q29), while 98% of studies had clearly defined, consistent data sources (Q30). Only 6% used large sample sizes (≥1000) for robust training and testing (Q31). Imaging protocols were standardized in 82% of studies (Q32), and 84% used automated segmentation (Q33). The impact of preprocessing (e.g., filtering, noise reduction) was discussed in 68% of studies (Q54).

***Preprocessing and Feature Engineering.*** Preprocessing (e.g., normalization, resampling) was consistently applied in 90% of studies (Q35), and 95% implemented measures to prevent data leakage (Q36). Dimension reduction was conducted independently of model training in 56% (Q37). Clinical relevance guided feature selection in 84% (Q38), and objective hyperparameter optimization was reported in 85% (Q39). Regularization (e.g., dropout, L2) was used in 66% of deep learning models to mitigate overfitting (Q40), which was observed in 36% of studies (Q41). Feature selection was justified contextually in 87% (Q42). Radiomics features were standardized using IBSI protocols in 48% (Q43), and preprocessing pipelines were fully disclosed in 91% of studies (Q44). Hyperparameter transparency was provided in 87% (Q45).

***Model Comparisons and Statistical Evaluation.*** Comparisons were made between deep radiomics and handcrafted radiomics in 10% of studies (Q46), deep radiomics vs. deep learning in4% (Q47), and handcrafted radiomics vs. deep learning in 9% (Q48). A full comparison among all three approaches (HRF, DRF, DL) was performed in 3% of studies (Q49). Statistical significance testing of model comparisons was done in 58% (Q50), and 85% assessed performance differences using statistical tests (Q51).

***Validation and Reproducibility Protocols.*** External testing using independent datasets was conducted in 71% of studies (Q52). Consistent missing data handling across models was reported in 29% (Q53). Reproducibility was supported in 88% of studies through the disclosure of parameters, software, and hardware used (Q54). Cross-validation procedures (e.g., k-fold, leave-one-out) were consistently applied across all model types in 84% of studies (Q55), and 88% used stratification or N-fold validation to ensure representative sampling (Q56).

***Transfer Learning.*** In deep radiomics studies, 26% used pre-trained networks for feature extraction (Q57), while 34% trained models from scratch and extracted features afterward (Q58). Transfer learning approaches were used and validated in 27% of deep learning models (Q59), indicating growing adoption of knowledge reuse from large-scale data.

## 4. Discussion

This systematic review addresses a critical and timely question in the rapidly evolving field of medical imaging: How do DL vs. machine learning techniques invoking HRFs and/or DRFs compare in predicting cancer outcomes using PET and SPECT imaging? Given the explosive growth of AI in oncology, clinicians face an urgent question: Which approach—DL, DRF, HRF, or fusion (a mixture of DRF, HRF, and clinical data)—is best suited for reliable, scalable cancer outcome prediction across imaging modalities? This review provides answers by systematically comparing these approaches across 231 studies, spanning a wide range of cancers—lung, breast, colorectal, prostate, and head and neck, and others—while also evaluating methodological rigor, clinical interpretability, and scalability. These insights are essential for selecting and developing AI models tailored to real-world clinical scenarios, especially in low-data environments where robust tools are most needed.

Our analysis revealed strong methodological rigor in the reviewed studies. Specifically, 96% employed proper dataset splitting into training, validation, and test sets; 84% used cross-validation techniques; and 95% conducted systematic model comparisons. Furthermore, 82% reported at least six evaluation metrics to ensure comprehensive assessments. Preprocessing methods were consistently applied in 90% of studies, and a notable proportion used multicenter data for training (67%) and external validation (71%). Outcome definitions were standardized in 91% of studies, and 97% ensured transparent performance comparisons. Despite these strengths, several methodological gaps were identified. Class imbalance was addressed in only 59% of studies. Socioeconomic and geographic variables were considered in just 9%, and only 19% acknowledged population diversity. Missing data handling was reported in 29%, and only 48% adopted IBSI-compliant radiomics standards—limiting reproducibility. Additionally, 59% of studies used small datasets, and only 55% employed multi-reader annotations for ground truth labeling—potentially compromising robustness.

Clinical translation remains hindered by these limitations. The opaque, "black-box" nature of DL and DRF models challenges integration into clinical workflows where interpretability may be critical (Bradshaw, McCradden et al. 2023). Despite their high performance, these models often lack transparency, which undermines clinician trust. To facilitate clinical adoption, strategies must include embedding explainable AI (XAI), standardizing preprocessing, improving population diversity, and adopting robust annotation practices. Techniques like SMOTE and focal loss should be applied to address class imbalance, and demographic variables must be incorporated to enhance fairness. Large, diverse datasets from multicenter collaborations are essential for building generalizable models. Regularization methods like dropout and L2 should also be consistently used to mitigate overfitting.

This review shows that DRF, fusion, and DL models consistently outperform HRF across PET and SPECT imaging for various cancer types. DRF demonstrated strong performance in data-limited settings, offering a practical alternative to DL due to its ability to extract complex, non-linear patterns with less data and computational demand. In smaller datasets, DRF outperformed HRF and, in many cases, DL. In larger datasets, DL models often surpassed DRF, but differences were not consistently statistically significant, highlighting the robustness of DRF. Fusion models maintained strong and stable performance across different dataset sizes, suggesting that combining DRF and HRF yields complementary benefits. Importantly, DRF excelled in limited-data scenarios, making it highly applicable in real-world clinical settings where large datasets are often unavailable. However, like DL, DRF models face challenges in interpretability, which remains a barrier to clinical implementation. Given the growing utility of DRF and the lack of a standardized extraction framework, future work should focus on developing DRF-specific pipelines. Such efforts would improve reproducibility, optimize predictive performance, and support broader clinical adoption across imaging modalities and cancer types.

Dataset size varied substantially across modeling approaches. HRF studies were the most prevalent (n = 127) and covered the broadest range of sample sizes (23 to 3,794 subjects) but were skewed toward smaller cohorts (median = 156; inter-quartile range (IQR) ≈ 100–251), reflecting the lower data and resource demands of handcrafted pipelines. DRF studies were relatively few (n = 20), typically centered around moderate sample sizes (median = 234; IQR ≈ 163–380). DL studies (n = 46) exhibited the widest upper tail, ranging from 26 to 15,000 subjects, with a long-tailed distribution (median = 244; IQR ≈ 158–682) shaped by a few large-scale datasets. Fusion models (n = 92) occupied an intermediate space, with sample sizes ranging from 43 to over 15,000; the median was 194 and IQR ≈ 118–336, showing moderate use of large datasets, especially when sourced from multi-center studies. These trends suggest that HRF remains the most accessible approach for small datasets, while DRF models are applied across an approximately similar sample size range. In contrast, DL models increasingly benefit from larger datasets, which may contribute to their superior performance in our accuracy and AUC analyses. DRF, both as standalone models and within fusion frameworks, demonstrated high point-estimate accuracy despite being underrepresented in the literature. Their ability to learn multiscale, non-linear patterns directly from imaging data, without manual ROI delineation or feature engineering, offers clear advantages in predictive modeling.

However, the limited number and smaller size of DRF alone and fusion studies result in wider confidence intervals and reduced statistical certainty. Furthermore, heterogeneity in model configurations, imaging protocols, and clinical populations was not controlled for, which may affect generalizability. As Vial et al. (Vial, Stirling et al. 2018) highlight, DL and DRF models typically require large annotated datasets—often thousands of cases—to avoid overfitting, while many HRF studies rely on a few hundred patients, often subdividing scans into patches to compensate. To confirm the potential of DRF models and mitigate overfitting risks, future studies should prioritize larger, multi-center datasets and adopt standardized validation frameworks that enable fair, reproducible comparisons across modeling strategies.

In PET imaging, DL and DRF (alone or mixed with HRF, i.e., fusion) models showed superior performance in predicting outcomes across lung, breast, prostate, and head and neck cancers. Hybrid models that integrated HRFs, DRFs, and clinical features frequently achieved AUCs exceeding 0.90 in treatment response and survival prediction. Studies by Ju et al. (Ju, Yang et al. 2023), Salmanpour et al. (Salmanpour, Rezaeijo et al. 2023), and Li et al. (Li, Han et al. 2024) demonstrated that fusion models combining clinical data with both HRFs and DRFs yielded the most generalizable outcomes. While HRFs offer strong interpretability, especially in harmonized multicenter datasets, their predictive power is generally surpassed by DL and DRF-based approaches. In SPECT imaging, DL models adapted well to low-resolution, high-noise data, outperforming HRFs in several cancers, including bone and thyroid malignancies. However, interpretability challenges across both DL and DRF methods continue to impede clinical uptake. Standardized imaging protocols and advanced denoising techniques may unlock greater predictive potential for SPECT in AI-based cancer modeling.

This review positions DRFs as a valuable compromise between feasibility and performance. They outperformed HRFs in small datasets and are more accessible than full DL pipelines, which demand greater data and computational resources. Nevertheless, like DL, DRFs lack interpretability, which restricts their clinical translation. Therefore, developing interpretable DRF pipelines and hybrid frameworks is crucial for real-world integration. PET remains the dominant modality, featured in 95.2% of studies, due to its superior spatial and temporal resolution that facilitates advanced feature extraction. While SPECT is underrepresented, it remains valuable in specific cancers like bone and thyroid—especially when enhanced with DL or hybrid models. To support wider adoption of SPECT, future work must focus on generating diverse datasets, harmonizing images, and incorporating clinical metadata. By addressing current methodological gaps and building on the predictive strengths of DRF and DL models, the clinical translation of PET and SPECT imaging in oncology becomes increasingly attainable. These advancements offer a path toward developing robust, personalized, and interpretable AI tools for cancer outcome prediction.

This review was evaluated by two nuclear medicine physicians, four medical physicists, and one radiology expert, all with expertise in DL and radiomics frameworks. The physicians appreciated the study's strong clinical relevance, particularly its focus on DRF and fusion models, which they believe are well-suited for outcome prediction in PET and SPECT, especially when data is limited. One physician offered a detailed critique of HRFs, emphasizing their rigidity and susceptibility to variations in imaging acquisition conditions—such as patient body habitus—even when the same scanner is used. In contrast, they highlighted that DRFs exhibit greater adaptability, making them more robust in real-world clinical scenarios. The ability of DRFs to extract and integrate information from multiple metastatic lesions, rather than focusing solely on a single dominant lesion as is common in HRF-based approaches, was noted as a major advantage. This broader sampling enables better representation of tumor heterogeneity and reflects the complex biological behavior of metastatic cancers. The physician also stressed the importance of developing standardized frameworks for DRFs, akin to the IBSI guidelines used for HRFs, to enhance their reproducibility and interpretability across clinical settings. The radiology expert highlighted the practical insights the study offers into AI integration within imaging pipelines and stressed the importance of addressing interpretability challenges to ensure clinical usability. They believe the study serves as a useful foundation for guiding radiology teams in selecting AI strategies based on available data and clinical needs. The medical physicists commended the methodological rigor and comparative approach, especially the evaluation of model performance across dataset sizes. They believe the study underscores the potential of DRFs as a bridge between handcrafted and deep learning methods. However, they raised concerns about the limited number of DRF studies and the variability in imaging protocols, which could affect reproducibility and generalizability. Collectively, these experts believe the study provides valuable, actionable guidance for AI model selection in oncology imaging and should inform future efforts to develop standardized, interpretable, and scalable DRF-based frameworks across clinical settings.

## 5. Conclusion

This systematic review of 231 studies demonstrates that DL and DRF-based models—whether used alone or in combination with HRFs (i.e., fusion models)—consistently outperform HRF-only approaches in cancer outcome prediction using PET and SPECT imaging. DRF-based models, particularly when used alone or within fusion frameworks, show strong performance in data-limited settings, offering an effective balance between predictive accuracy and computational efficiency. While DL models excel in large datasets, DRFs remain highly competitive in moderate-sized cohorts and are especially valuable where annotated data are scarce. Fusion models further boost performance by leveraging complementary information from both DRFs and HRFs. However, both DL and DRF-based models face challenges related to limited interpretability. In addition, the limited number of studies focusing on DRFs and the inconsistent modeling approaches used across these studies make it difficult to draw general conclusions or apply the findings broadly. Future research should prioritize the development of standardized and interpretable DRF extraction frameworks. In particular, building foundational tools for consistent DRF extraction across imaging modalities and disease types will be essential for enabling robust, reproducible, and clinically meaningful AI applications in oncology.

**Acknowledgment.** This study was supported by the Virtual Collaboration Group (VirCollab, www.vircollab.com) and the Technological Virtual Collaboration (TECVICO CORP.) based in Vancouver, Canada. We gratefully acknowledge funding from the Canadian Foundation for Innovation – John R. Evans Leaders Fund (CFI-JELF; Award No. AWD-023869 CFI), as well as the Natural Sciences and Engineering Research Council of Canada (NSERC) Awards AWD-024385, RGPIN-2023-0357, and Discovery Horizons Grant DH-2025-00119.

**Conflict of Interest**. The co-authors Sonya Falahati, and Mohammad R. Salmanpour, are affiliated with TECVICO CORP. The remaining co-authors declare no relevant conflicts of interest.

**Supplemental Files**. Supplemental Files 1 and 2 are available at the links provided below.
*https://drive.google.com/drive/folders/1OeC21QA5q3_6pUTNnLtDpYdIXghJxyYm?usp=sharing*

## References

Abdallah, N., J. M. Marion, C. Tauber, T. Carlier, P. Chauvet and M. Hatt (2023). Deep Learning-Based Outcome Prediction Using Harmonized CT Images for Head and Neck Cancer Patients: A Novel Approach with Attention and DeepHitSingle. 2023 IEEE Nuclear Science Symposium, Medical Imaging Conference and International Symposium on Room-Temperature Semiconductor Detectors (NSS MIC RTSD).

Ahmed, S. F., M. S. B. Alam, M. Hassan, M. R. Rozbu, T. Ishtiak, N. Rafa, M. Mofijur, A. B. M. Shawkat Ali and A. H. Gandomi (2023). "Deep learning modelling techniques: current progress, applications, advantages, and challenges." Artificial Intelligence Review 56(11): 13521–13617.

Ahrari, S., T. Zaragori, A. Zinsz, J. Oster, L. Imbert and A. Verger (2024). "Application of PET imaging delta radiomics for predicting progression-free survival in rare high-grade glioma." Scientific Reports 14(1): 3256.

Aksu, A., Z. G. Güç, K. A. Küçüker, A. Alacacıoğlu and B. Turgut (2024). "Intra and peritumoral PET radiomics analysis to predict the pathological response in breast cancer patients receiving neoadjuvant chemotherapy." Revista Española de Medicina Nuclear e Imagen Molecular (English Edition) 43(3): 500002.

Al-Battat, K. M. (2024). Prediction of Mediastinal Lymph Node Metastasis in Non-Small Cell Lung Cancer Based on 18F-FDG PET Radiomics. 2024 IEEE Nuclear Science Symposium (NSS), Medical Imaging Conference (MIC) and Room Temperature Semiconductor Detector Conference (RTSD).

Alqahtani, F. F. (2023). "SPECT/CT and PET/CT, related radiopharmaceuticals, and areas of application and comparison." Saudi Pharmaceutical Journal 31(2): 312–328.

Amini, M., M. Nazari, I. Shiri, G. Hajianfar, M. R. Deevband, H. Abdollahi and H. Zaidi (2020). Multi-Level PET and CT Fusion Radiomics-based Survival Analysis of NSCLC Patients. 2020 IEEE Nuclear Science Symposium and Medical Imaging Conference (NSS/MIC).

Andrearczyk, V., P. Fontaine, V. Oreiller, J. Castelli, M. Jreige, J. O. Prior and A. Depeursinge (2021). Multi-task Deep Segmentation and Radiomics for Automatic Prognosis in Head and Neck Cancer. 4th International Workshop on Predictive Intelligence in Medicine, PRIME 2021, held in conjunction with 24th International Conference on Medical Image Computing and Computer Assisted Intervention, MICCAI 2021, Strasbourg, France, Springer Science and Business Media Deutschland GmbH.

Andrearczyk, V., V. Oreiller, S. Boughdad, C. C. Le Rest, O. Tankyevych, H. Elhalawani, M. Jreige, J. O. Prior, M. Vallières, D. Visvikis, M. Hatt and A. Depeursinge (2023). "Automatic Head and Neck Tumor segmentation and outcome prediction relying on FDG-PET/CT images: Findings from the second edition of the HECKTOR challenge." Medical Image Analysis 90: 102972.

Andrew William, C., P. Ohm, A. C. Eric, R. Leonid and K. Despina (2024). Radiomic phenotypes of the background lung parenchyma from [18]F-FDG PET/CT images can augment tumor radiomics and clinical factors in predicting response after surgical resection of tumors in patients with non-small cell lung cancer. Proc.SPIE.

Arabi, H., A. AkhavanAllaf, A. Sanaat, I. Shiri and H. Zaidi (2021). "The promise of artificial intelligence and deep learning in PET and SPECT imaging." Physica Medica 83: 122–137.

Arabi, H., I. Shiri, E. Jenabi, M. Becker and H. Zaidi (2020). Deep Learning-based Automated Delineation of Head and Neck Malignant Lesions from PET Images. 2020 IEEE Nuclear Science Symposium and Medical Imaging Conference (NSS/MIC).

Avval, A. H., M. Amini, G. Hajianfar, I. Shiri and H. Zaidi (2022). Progression-Free Survival Prediction in Head and Neck Cancer using Fused PET-CT Radiomics and Machine Learning. 2022 IEEE Nuclear Science Symposium and Medical Imaging Conference (NSS/MIC).

Barlow, S. H., S. Chicklore, Y. He, S. Ourselin, T. Wagner, A. Barnes and G. J. R. Cook (2024). "Uncertainty-aware automatic TNM staging classification for [18F] Fluorodeoxyglucose PET-CT reports for lung cancer utilising transformer-based language models and multi-task learning." BMC Medical Informatics and Decision Making 24(1): 396.

Bauckneht, M., G. Pasini, T. Di Raimondo, G. Russo, S. Raffa, M. I. Donegani, D. Dubois, L. Peñuela, L. Sofia, G. Celesti, F. Bini, F. Marinozzi, F. Lanfranchi, R. Laudicella, G. Sambuceti and A. Stefano (2025). "[(18)F]PSMA-1007 PET/CT-based radiomics may help enhance the interpretation of bone focal uptakes in hormone-sensitive prostate cancer patients." Eur J Nucl Med Mol Imaging.

Behmanesh, B., A. Abdi-Saray, M. R. Deevband, M. Amoui, H. R. Haghighatkhah and A. Shalbaf (2024). "Predicting the Response of Patients Treated with (177)Lu-DOTATATE Using Single-photon Emission Computed Tomography-Computed Tomography Image-based Radiomics and Clinical Features." J Med Signals Sens 14: 28.

Berti, V., E. Fasciglione, A. Charpiot, F. Montanini, M. Pepponi, A. Leo, F. Hubele, D. Taieb, K. Pacak, B. Goichot and A. Imperiale (2025). "Deciphering 18F-DOPA uptake in SDH-related head and neck paragangliomas: a radiomics approach." Journal of Endocrinological Investigation 48(4): 941–950.

Beukinga, R. J., F. B. Poelmann, G. Kats-Ugurlu, A. R. Viddeleer, R. Boellaard, R. J. De Haas, J. T. M. Plukker and J. B. Hulshoff (2022) "Prediction of Non-Response to Neoadjuvant Chemoradiotherapy in Esophageal Cancer Patients with 18F-FDG PET Radiomics Based Machine Learning Classification." Diagnostics 12 DOI: 10.3390/diagnostics12051070.

Bian, S., W. Hong, X. Su, F. Yao, Y. Yuan, Y. Zhang, J. Xie, T. Li, K. Pan, Y. Xue, Q. Zhang, Z. Yu, K. Tang, Y. Yang, Y. Zhu ang, J. Lin and H. Xu (2024). "A dynamic online nomogram predicting prostate cancer short-term prognosis based on (18)F-PSMA-1007 PET/CT of periprostatic adipose tissue: a multicenter study." Abdom Radiol (NY) 49(10): 3747–3757.

Bianconi, F., R. Salis, M. L. Fravolini, M. U. Khan, L. Filippi, A. Marongiu, S. Nuvoli, A. Spanu and B. Palumbo (2024). "Rad iomics Features from Positron Emission Tomography with [(18)F] Fluorodeoxyglucose Can Help Predict Cervical Nodal Status in Patients with Head and Neck Cancer." Cancers (Basel) 16(22).

Bicakci, M., O. Ayyildiz, Z. Aydin, A. Basturk, S. Karacavus and B. Yilmaz (2020). "Metabolic Imaging Based Sub -Classification of Lung Cancer." IEEE Access 8: 218470–218476.

Bradshaw, T. J., M. D. McCradden, A. K. Jha, J. Dutta, B. Saboury, E. L. Siegel and A. Rahmim (2023). "Artificial Intelligenc e Algorithms Need to Be Explainable-or Do They?" J Nucl Med 64(6): 976–977.

Breylon, A. R., B. S. Jack, L. Xiang, Y. Zhenyu, W. Chunhao, M. Yvonne Marie, M. B. David, Y. Fang-Fang and J. L. Kyle (2024). "Prognostic value of different discretization parameters in <sup>18</sup>fluorodeoxyglucose positron emission tomography radiomics of oropharyngeal squamous cell carcinoma." Journal of Medical Imaging 11(2): 024007.

Cepero, A., Y. Yang, L. Young, J. Huang, X. Ji and F. Yang (2024) "Longitudinal FDG-PET Radiomics for Early Prediction of Treatment Response to Chemoradiation in Locally Advanced Cervical Cancer: A Pilot Study." Cancers 16 DOI: 10.3390/cancers16223813.

Cepero, A. A. Y., Y.; Young, L.; Yang, F. (2024). "Predictive Value of FDG PET Radiomics for Early Response to Chemoradiation in Locally Advanced Cervical Cancer, in Proceedings of the 4th International Electronic Conference on Cancers."

Chang, W., S. Zhou, D. Sun, Y. Liu, W. Mao, W. Cen, W. Tang, L. Ye, L. Wang and J. Xu (2022). "53P Baseline PET/CT deep radiomics signature apply for identifying bevacizumab sensitivity of RAS-mutant colorectal cancer liver metastases patients." Annals of Oncology 33: S1449.

Chen, D., R. Zhou and B. Li (2024). "Preoperative Prediction of Her-2 and Ki-67 Status in Gastric Cancer Using (18)F-FDG PET/CT Radiomics Features of Visceral Adipose Tissue." Br J Hosp Med (Lond) 85(9): 1–18.

Chen, L., Y. Luo and Q. Wang (2021). "Radiomics signature: a potential biomarker for the prediction of MCI conversion to Alzheimer's disease." Front Aging Neurosci 13: 645462.

Cheng, L., H. Gao, Z. Wang, L. Guo, X. Wang and G. Jin (2024). "Prospective study of dual-phase (99m)Tc-MIBI SPECT/CT nomogram for differentiating non-small cell lung cancer from benign pulmonary lesions." Eur J Radiol 179: 111657.

Cheng, Z., P. Chen and J. Yan (2025). "A review of state-of-the-art resolution improvement techniques in SPECT imaging." EJNMMI Physics 12(1): 9.

Ciarmiello, A., E. Giovannini, F. Tutino, N. Yosifov, A. Milano, L. Florimonte, E. Bonatto, C. Bareggi, L. Dellavedova, A. Ca stello, C. Aschele, M. Castellani and G. Giovacchini (2024). "Does FDG PET-Based Radiomics Have an Added Value for Prediction of Overall Survival in Non-Small Cell Lung Cancer?" J Clin Med 13(9).

Collarino, A., V. Feudo, T. Pasciuto, A. Florit, E. Pfaehler, M. de Summa, N. Bizzarri, S. Annunziata, G. F. Zannoni, L. F. d e Geus-Oei, G. Ferrandina, M. A. Gambacorta, G. Scambia, R. Boellaard, E. Sala, V. Rufini and F. H. van Velden (2024). "Is PET Radi omics Useful to Predict Pathologic Tumor Response and Prognosis in Locally Advanced Cervical Cancer?" J Nucl Med 65(6): 962–970.

Crişan, G., N. S. Moldovean-Cioroianu, D. G. Timaru, G. Andrieş, C. Căinap and V. Chiş (2022). "Radiopharmaceuticals for PET and SPECT Imaging: A Literature Review over the Last Decade." Int J Mol Sci 23(9).

Czernin, J., M. Allen-Auerbach, D. Nathanson and K. Herrmann (2013). "PET/CT in Oncology: Current Status and Perspectives." Current Radiology Reports 1(3): 177–190.

Da-Ano, R., O. Tankyevych, G. Andrade-Miranda, P. H. Conze, C. C. L. Rest and D. Visvikis (2024). Multi-Modal PET/CT Fusion for Automated PD-L1 Status Prediction in Lung Cancer. 2024 IEEE International Symposium on Biomedical Imaging (ISBI).

de Geus-Oei, L. F., D. Vriens, H. W. van Laarhoven, W. T. van der Graaf and W. J. Oyen (2009). "Monitoring and predicting response to therapy with 18F-FDG PET in colorectal cancer: a systematic review." J Nucl Med 50 Suppl 1: 43s–54s.

Dholey, M., R. J. M. Santosham, S. Ray, J. Das, S. Chatterjee, R. Ahmed and J. Mukherjee (2023). Ensemble Methods with [18F]FDG-PET/CT Radiomics in Breast Cancer Response Prediction. Pattern Recognition and Machine Intelligence: 10th International Conference, PReMI 2023, Kolkata, India, December 12–15, 2023, Proceedings. Kolkata, India, Springer-Verlag: 369–379.

Diao, Z. and H. Jiang (2024). "A multi-instance tumor subtype classification method for small PET datasets using RA-DL attention module guided deep feature extraction with radiomics features." Comput Biol Med 174: 108461.

Ding, H., Y. Li, T. Liang, Y. Liao, X. Yu, X. Duan and C. Shen (2023). PET/CT Radiomics Integrated with Clinical Indexes as A Tool to Predict Ki67 in Breast Cancer: A Pilot Study.

Duan, F., M. Zhang, C. Yang, X. Wang and D. Wang (2025). "Non-invasive Prediction of Lymph Node Metastasis in NSCLC Using Clinical, Radiomics, and Deep Learning Features From (18)F-FDG PET/CT Based on Interpretable Machine Learning." Acad Radiol 32(3): 1645–1655.

Dwarakadeesh, K., S. Shinde and P. Kanagaraju (2024). Multimodal Image Fusion for Enhanced Brain Tumor Detection using Advanced Machine Learning Techniques. 2024 Second International Conference on Inventive Computing and Informatics (ICICI).

Eifer, M., G. Peters-Founshtein, L. C. Yoel, H. Pinian, R. Steiner, E. Klang, O. A. Catalano, Y. Eshet and L. Domachevsky (2024). "The role of FDG PET/CT radiomics in the prediction of pathological response to neoadjuvant treatment in patients with esophageal cancer." Rep Pract Oncol Radiother 29(2): 211–218.

Eifer, M., H. Pinian, E. Klang, Y. Alhoubani, N. Kanana, N. Tau, T. Davidson, E. Konen, O. A. Catalano, Y. Eshet and L. Domachevsky (2022). "FDG PET/CT radiomics as a tool to differentiate between reactive axillary lymphadenopathy following COVID-19 vaccination and metastatic breast cancer axillary lymphadenopathy: a pilot study." Eur Radiol 32(9): 5921–5929.
Fan, R., X. Long, X. Chen, Y. Wang, D. Chen and R. Zhou (2025). "The Value of Machine Learning-based Radiomics Model Characterized by PET Imaging with (68)Ga-FAPI in Assessing Microvascular Invasion of Hepatocellular Carcinoma." Acad Radiol 32(4): 2233–2246.
Fan, X., H. Zhang, Z. Wang, X. Zhang, S. Qin, J. Zhang, F. Hu, M. Yang, J. Zhang and F. Yu (2024). "Diagnosing postoperative lymph node metastasis in thyroid cancer with multimodal radiomics and clinical features." Digit Health 10: 20552076241233244.
Faraji, S., Emami, F., Vosoughi, Z. et al. . J. Med. Biol. Eng. 44, (2024) (2024). "Predicting Immunohistochemical Biomarkers of Breast Cancer Using 18F-FDG PET/CT Radiomics: A Multicenter Study." J. Med. Biol. Eng 44: 749–762.
Fathi Jouzdani, A., A. Abootorabi, M. Rajabi, F. Panahabadi, A. Gorji, N. Sanati, A. M. Ahmadzadeh, A. Mousavi, L. Bonnie, C. Ho, R. Yuan, A. Rahmim and M. R. Salmanpour (2024). "<strong>Impact of Clinical Features Combined with PET/CT Imaging Features on Survival Prediction of Outcome in Lung Cancer </strong>." Journal of Nuclear Medicine 65(supplement 2): 242130.
Feng, L., X. Yang, X. Lu, Y. Kan, C. Wang, D. Sun, H. Zhang, W. Wang and J. Yang (2022). "(18)F-FDG PET/CT-based radiomics nomogram could predict bone marrow involvement in pediatric neuroblastoma." Insights Imaging 13(1): 144.
Feng, L., X. Yang, C. Wang, H. Zhang, W. Wang and J. Yang (2024). "Predicting event-free survival after induction of remission in high-risk pediatric neuroblastoma: combining (123)I-MIBG SPECT-CT radiomics and clinical factors." Pediatr Radiol 54(5): 805–819.
Feng, L., X. Yao, X. Lu, C. Wang, W. Wang and J. Yang (2024). "Differentiation of early relapse and late relapse in intermediate- and high-risk neuroblastoma with an (18)F-FDG PET/CT-based radiomics nomogram." Abdom Radiol (NY) 49(3): 888–899.
Feng, L., X. Yao, C. Wang, H. Zhang, W. Wang and J. Yang (2025). "Radiomics analysis based on 18F-fluorodeoxyglucose positron emission tomography/computed tomography for differentiating the histological classification of peripheral neuroblastic tumours." Clinical Radiology 84: 106851.
Feng, L., Z. Zhou, J. Liu, S. Yao, C. Wang, H. Zhang, P. Xiong, W. Wang and J. Yang (2024). "18F-FDG PET/CT-Based Radiomics Nomogram for Prediction of Bone Marrow Involvement in Pediatric Neuroblastoma: A Two-Center Study." Academic Radiology 31(3): 1111–1121.
Fiz, F., C. Masci, G. Costa, M. Sollini, A. Chiti, F. Ieva, G. Torzilli and L. Viganò (2022). "PET/CT-based radiomics of mass-forming intrahepatic cholangiocarcinoma improves prediction of pathology data and survival." Eur J Nucl Med Mol Imaging 49(10): 3387–3400.
Fujarewicz, K., A. Wilk, D. Borys, A. d'Amico, R. Suwiński and A. Świerniak (2022). Machine Learning Approach to Predict Metastasis in Lung Cancer Based on Radiomic Features. Intelligent Information and Database Systems, Cham, Springer Nature Switzerland.
Fujima, N., V. C. Andreu-Arasa, S. K. Meibom, G. A. Mercier, M. T. Truong, K. Hirata, K. Yasuda, S. Kano, A. Homma, K. Kudo and O. Sakai (2021). "Prediction of the local treatment outcome in patients with oropharyngeal squamous cell carcinoma using deep learning analysis of pretreatment FDG-PET images." BMC Cancer 21(1): 900.
Furukawa, M., D. R. McGowan and B. W. Papież (2024). Prediction of Recurrence Free Survival of Head and Neck Cancer Using PET/CT Radiomics and Clinical Information. 2024 IEEE International Symposium on Biomedical Imaging (ISBI).
Gatenby, R. A., O. Grove and R. J. Gillies (2013). "Quantitative Imaging in Cancer Evolution and Ecology." Radiology 269(1): 8–14.
Gelardi, F., L. Cavinato, R. de Sanctis, G. Ninatti, P. Tiberio, M. Rodari, A. Zambelli, A. Santoro, B. Fernandes, A. Chiti, L. Antunovic and M. Sollini (2024). "The Predictive Role of Radiomics in Breast Cancer Patients Imaged by [18F]FDG PET: Preliminary Results from a Prospective Cohort." Diagnostics 14.
Gil, J., H. Choi, J. C. Paeng, G. J. Cheon and K. W. Kang (2023). "Deep Learning-Based Feature Extraction from Whole-Body PET/CT Employing Maximum Intensity Projection Images: Preliminary Results of Lung Cancer Data." Nuclear Medicine and Molecular Imaging 57(5): 216–222.
Gorji, A., M. Hosseinzadeh, A. F. Jouzdani, N. Sanati, F. Y. Rizi, S. Moore, B. Leung, C. Ho, I. Shiri, H. Zaidi, A. Rahmim and M. R. Salmanpour (2023). Region-of-Interest and Handcrafted vs. Deep Radiomics Feature Comparisons for Survival Outcome Prediction: Application to Lung PET/CT Imaging. 2023 IEEE Nuclear Science Symposium, Medical Imaging Conference and International Symposium on Room-Temperature Semiconductor Detectors (NSS MIC RTSD).
Gu, B., M. Meng, L. Bi, J. Kim, D. D. Feng and S. Song (2022). "Prediction of 5-year progression-free survival in advanced nasopharyngeal carcinoma with pretreatment PET/CT using multi-modality deep learning-based radiomics." Front Oncol 12: 899351.
Gutsche, R., J. Scheins, M. Kocher, K. Bousabarah, G. R. Fink, N. J. Shah, K. J. Langen, N. Galldiks and P. Lohmann (2021). "Evaluation of FET PET Radiomics Feature Repeatability in Glioma Patients." Cancers (Basel) 13(4).
Haider, S. P., A. Mahajan, T. Zeevi, P. Baumeister, C. Reichel, K. Sharaf, R. Forghani, A. S. Kucukkaya, B. H. Kann, B. L. Judson, M. L. Prasad, B. Burtness and S. Payabvash (2020). "PET/CT radiomics signature of human papilloma virus association in oropharyngeal squamous cell carcinoma." Eur J Nucl Med Mol Imaging 47(13): 2978–2991.
Han, Y., G. Wang, J. Zhang, Y. Pan, J. Cui, C. Li, Y. Wang, X. Xu and B. Xu (2024). "The value of radiomics based on 2-[18 F]FDG PET/CT in predicting WHO/ISUP grade of clear cell renal cell carcinoma." EJNMMI Res 14(1): 115.
Hatt, M., A. K. Krizsan, A. Rahmim, T. J. Bradshaw, P. F. Costa, A. Forgacs, R. Seifert, A. Zwanenburg, I. El Naqa, P. E. Kinahan, F. Tixier, A. K. Jha and D. Visvikis (2023). "Joint EANM/SNMMI guideline on radiomics in nuclear medicine : Jointly supported by the EANM Physics Committee and the SNMMI Physics, Instrumentation and Data Sciences Council." Eur J Nucl Med Mol Imaging 50(2): 352–375.
Hinzpeter, R., S. A. Mirshahvalad, V. Murad, L. Avery, R. Kulanthaivelu, A. Kohan, C. Ortega, E. Elimova, J. Yeung, A. Hope, U. Metser and P. Veit-Haibach (2024). "The [(18)F]F-FDG PET/CT Radiomics Classifier of Histologic Subtypes and Anatomical Disease Origins across Various Malignancies: A Proof-of-Principle Study." Cancers (Basel) 16(10).
Holzschuh, J. C., M. Mix, J. Ruf, T. Hölscher, J. Kotzerke, A. Vrachimis, P. Doolan, H. Ilhan, I. M. Marinescu, S. K. B. Spohn, T. Fechter, D. Kuhn, P. Bronsert, C. Gratzke, R. Grosu, S. C. Kamran, P. Heidari, T. S. C. Ng, A. Könik, A.-L. Grosu and C. Zamboglou (2023). "Deep learning based automated delineation of the intraprostatic gross tumour volume in PSMA-PET for patients with primary prostate cancer." Radiotherapy and Oncology 188: 109774.
Hossain, T., Z. Qureshi, N. Jayakumar, T. E. Muttikkal, S. Patel, D. Schiff, M. Zhang and B. Kundu (2023). Multimodal Deep Learning to Differentiate Tumor Recurrence from Treatment Effect in Human Glioblastoma. 2023 IEEE 20th International Symposium on Biomedical Imaging (ISBI).
Hosseini, S. A., G. Hajianfar, P. Ghaffarian, M. Seyfi, E. Hosseini, A. H. Aval, S. Servaes, M. Hanaoka, P. Rosa-Neto, S. Chawla, H. Zaidi and M. R. Ay (2024). "PET radiomics-based lymphovascular invasion prediction in lung cancer using multiple segmentation and multi-machine learning algorithms." Phys Eng Sci Med 47(4): 1613–1625.
Hosseini, S. A., G. Hajianfar, I. Shiri and H. Zaidi (2021). Lung Cancer Recurrence Prediction Using Radiomics Features of PET Tumor Sub-Volumes and Multi-Machine Learning Algorithms. 2021 IEEE Nuclear Science Symposium and Medical Imaging Conference (NSS/MIC).
Hou, X., K. Chen, X. Wan, H. Luo, X. Li and W. Xu (2024). "Intratumoral and peritumoral radiomics for preoperative prediction of neoadjuvant chemotherapy effect in breast cancer based on (18)F-FDG PET/CT." J Cancer Res Clin Oncol 150(11): 484.
Huang, B., Q. Yang, X. Li, Y. Wu, Z. Liu, Z. Pan, S. Zhong, S. Song and C. Zuo (2024). "Deep learning–based whole-body characterization of prostate cancer lesions on [68Ga]Ga-PSMA-11 PET/CT in patients with post-prostatectomy recurrence." European Journal of Nuclear Medicine and Molecular Imaging 51(4): 1173–1184.
Huang, J., T. Li, L. Tang, Y. Hu, Y. Hu and Y. Gu (2024). "Development and Validation of an 18F-FDG PET/CT-based Radiomics Nomogram for Predicting the Prognosis of Patients with Esophageal Squamous Cell Carcinoma." Academic Radiology 31(12): 5066–5077.
Huang, S., C. Cao, L. Guo, C. Li, F. Zhang, Y. Li, Y. Liang and W. Mu (2024). "Comparison of the variability and diagnostic efficacy of respiratory-gated PET/CT based radiomics features with ungated PET/CT in lung lesions." Lung Cancer 194: 107889.
Huang, W., J. Wang, H. Wang, Y. Zhang, F. Zhao, K. Li, L. Su, F. Kang and X. Cao (2022). "PET/CT Based EGFR Mutation Status Classification of NSCLC Using Deep Learning Features and Radiomics Features." Frontiers in Pharmacology 13.
Huang, Z., Y. Zhu, W. Li, L. Yang, Y. Yang, D. Liang, D. Shao and Z. Hu (2023). Prediction of brain metastases from nonsmall cell lung cancer using multimodal radiomics and deep learning based on 18FPET/CT images. 2023 IEEE Nuclear Science Symposium, Medical Imaging Conference and International Symposium on Room-Temperature Semiconductor Detectors (NSS MIC RTSD).
Ingenerf, M., F. Grawe, M. Winkelmann, H. Karim, J. Ruebenthaler, M. P. Fabritius, J. Ricke, R. Seidensticker, C. J. Auernhammer, M. J. Zacherl, M. Seidensticker and C. Schmid-Tannwald (2024). "Neuroendocrine liver metastases treated using transarterial radioembolization: Identification of prognostic parameters at 68Ga-DOTATATE PET/CT." Diagnostic and Interventional Imaging 105(1): 15–25.
Inglese, M., A. Duggento, T. Boccato, M. Ferrante and N. Toschi (2022). Spatiotemporal learning of dynamic positron emission tomography data improves diagnostic accuracy in breast cancer. 2022 44th Annual International Conference of the IEEE Engineering in Medicine & Biology Society (EMBC).
Jia, L., Z. Chen, D. Shao and A. A. Abba (2023). Breast Cancer Subtype Classification Based on PET/CT Bimodal Imaging Feature Fusion. 2023 IEEE 6th International Conference on Pattern Recognition and Artificial Intelligence (PRAI).
Jimenez-Mesa, C., J. E. Arco, F. J. Martinez-Murcia, J. Suckling, J. Ramirez and J. M. Gorriz (2023). "Applications of machine learning and deep learning in SPECT and PET imaging: General overview, challenges and future prospects." Pharmacological Research 197: 106984.
Jin, Z., F. Zhang, Y. Wang, A. Tian, J. Zhang, M. Chen and J. Yu (2021). "Single-Photon Emission Computed Tomography/Computed Tomography Image-Based Radiomics for Discriminating Vertebral Bone Metastases From Benign Bone Lesions in Patients With Tumors." Front Med (Lausanne) 8: 792581.
Ju, H. M., J. Yang, J. M. Park, J. H. Choi, H. Song, B. I. Kim, U. S. Shin, S. M. Moon, S. Cho and S. K. Woo (2023). "Prediction of Neoadjuvant Chemoradiotherapy Response in Rectal Cancer Patients Using Harmonized Radiomics of Multcenter (18)F-FDG-PET Image." Cancers (Basel) 15(23).
Ju, L., W. Li, R. Zuo, Z. Chen, Y. Li, Y. Feng, Y. Xiang and H. Pang (2024). "Deep Learning Features and Metabolic Tumor Volume Based on PET/CT to Construct Risk Stratification in Non-small Cell Lung Cancer." Academic Radiology 31(11): 4661–4675.
Kaiser, L., S. Quach, A. J. Zounek, B. Wiestler, A. Zatcepin, A. Holzgreve, A. Bollenbacher, L. M. Bartos, V. C. Ruf, G. Böning, N. Thon, J. Herms, M. J. Riemenschneider, S. Stöcklein, M. Brendel, R. Rupprecht, J. C. Tonn, P. Bartenstein, L. von Baumgarten, S. Ziegler and N. L. Albert (2024). "Enhancing predictability of IDH mutation status in glioma patients at initial diagnosis: a comparative analysis of radiomics from MRI, [(18)F]FET PET, and TSPO PET." Eur J Nucl Med Mol Imaging 51(8): 2371–2381.
Kang, Y.-k., S. Ha, J. B. Jeong and S. W. Oh (2024). "The value of PET/CT radiomics for predicting survival outcomes in patients with pancreatic ductal adenocarcinoma." Scientific Reports 14(1): 28958.
Kawahara, D., R. Nishioka, Y. Murakami, Y. Emoto, K. Iwashita and R. Sasaki (2024). "A nomogram based on pretreatment radiomics and dosiomics features for predicting overall survival associated with esophageal squamous cell cancer." European Journal of Surgical Oncology 50(7): 108450.

Kawauchi, K., S. Furuya, K. Hirata, C. Katoh, O. Manabe, K. Kobayashi, S. Watanabe and T. Shiga (2020). "A convolutional neural network-based system to classify patients using FDG PET/CT examinations." BMC Cancer 20(1): 227.
Kelk, E., P. Ruuge, K. Rohtla, A. Poksi and K. Kairemo (2021). "Radiomics Analysis for (177)Lu-DOTAGA-(l-y)fk(Sub-KuE) Targeted Radioligand Therapy Dosimetry in Metastatic Prostate Cancer-A Model Based on Clinical Example." Life (Basel) 11(2).
Kim, C., H. H. Cho, J. Y. Choi, T. J. Franks, J. Han, Y. Choi, S. H. Lee, H. Park and K. S. Lee (2021). "Pleomorphic carcinoma of the lung: Prognostic models of semantic, radiomics and combined features from CT and PET/CT in 85 patients." Eur J Radiol Open 8: 100351.
Klontzas, M. E., E. Koltsakis, G. Kalarakis, K. Trpkov, T. Papathomas, A. H. Karantanas and A. Tzortzakakis (2023). "Machine Learning Integrating (99m)Tc Sestamibi SPECT/CT and Radiomics Data Achieves Optimal Characterization of Renal Oncocytic Tumors." Cancers (Basel) 15(14).
Kovacs, D. G., C. N. Ladefoged, K. F. Andersen, J. M. Brittain, C. B. Christensen, D. Dejanovic, N. L. Hansen, A. Loft, J. H. Petersen, M. Reichkendler, F. L. Andersen and B. M. Fischer (2024). "Clinical Evaluation of Deep Learning for Tumor Delineation on <sup>18</sup>F-FDG PET/CT of Head and Neck Cancer." Journal of Nuclear Medicine: jnumed.123.266574.
Kumar, N., S. A. Shamim, H. Gupta, A. Mehndiratta, G. Arora, E. Kayal and S. Jaswal (2024). "<strong>Prediction of diagnostic accuracy with 18F-FDG PET/CT using PET semi-quantitative and radiomics features in pre-nephrectomy Renal Cell Cancer (RCC) patients</strong>." Journal of Nuclear Medicine 65(supplement 2): 242041–242041.
Kumar, R., A. Ramachandran, B. R. Mittal and H. Singh (2024). "Convoluted Neural Network for Detection of Clinically Significant Prostate Cancer on 68 Ga PSMA PET/CT Delayed Imaging by Analyzing Radiomic Features." Nuclear Medicine and Molecular Imaging 58(2): 62–68.
Kun, H., M. Gao, E. Antonecchia, L. Zhang, Z. Zhou, X. Zou, Z. Li, W. Cao, Y. Liu and N. D'Ascenzo (2024). "Enhanced Risk Stratification of Gastrointestinal Stromal Tumors Through Cross-Modality Synthesis from CT to [$^{18}$F]-FDG PET Images." IEEE Transactions on Radiation and Plasma Medical Sciences PP: 1–1.
Kundu, B., W. Terrell, Z. Qureshi, R. Chavan, S. Patel, P. Batchala, S. Berr, S. Moosa, M. Quigg, B. Purow, T. Eluvathingal Muttikkal and D. Schiff (2024). "<strong>Dynamic FDG PET for multimodal classification of tumor progression from treatment effect in human glioblastoma: Regression, Radiomics and Deep learning</strong>." Journal of Nuclear Medicine 65(supplement 2): 241157–241157.
Kwon, R., H. Kim, K. S. Ahn, B.-I. Song, J. Lee, H. W. Kim, K. S. Won, H. W. Lee, T.-S. Kim, Y. Kim and K. J. Kang (2024) "A Machine Learning-Based Clustering Using Radiomics of F-18 Fluorodeoxyglucose Positron Emission Tomography/Computed Tomography for the Prediction of Prognosis in Patients with Intrahepatic Cholangiocarcinoma." Diagnostics 14 DOI: 10.3390/diagnostics14192245.
Lambin, P., E. Rios-Velazquez, R. Leijenaar, S. Carvalho, R. G. van Stiphout, P. Granton, C. M. Zegers, R. Gillies, R. Boellard, A. Dekker and H. J. Aerts (2012). "Radiomics: extracting more information from medical images using advanced feature analysis." Eur J Cancer 48(4): 441–446.
Lee, J., J. Lee and B.-I. Song (2025). "A Machine Learning-Based Radiomics Model for the Differential Diagnosis of Benign and Malignant Thyroid Nodules in F-18 FDG PET/CT: External Validation in the Different Scanner." Cancers 17(2): 331.
Lee, J. W. and S. M. Lee (2018). "Radiomics in Oncological PET/CT: Clinical Applications." Nuclear Medicine and Molecular Imaging 52(3): 170–189.
Leung, K. H., S. P. Rowe, M. S. Sadaghiani, J. P. Leal, E. Mena, P. L. Choyke, Y. Du and M. G. Pomper (2024). "Deep Semisupervised Transfer Learning for Fully Automated Whole-Body Tumor Quantification and Prognosis of Cancer on PET/CT." J Nucl Med 65(4): 643–650.
Li, B., J. Su, K. Liu and C. Hu (2024). "Deep learning radiomics model based on PET/CT predicts PD-L1 expression in non-small cell lung cancer." European Journal of Radiology Open 12: 100549.
Li, S., Y. Hu, C. Tian, J. Luan, X. Zhang, Q. Wei, X. Li and Y. Bian (2024). "Prediction of EGFR-TP53 genes co-mutations in patients with lung adenocarcinoma (LUAD) by 18F-FDG PET/CT radiomics." Clinical and Translational Oncology.
Li, T., J. Mao, J. Yu, Z. Zhao, M. Chen, Z. Yao, L. Fang and B. Hu (2024). "Fully automated classification of pulmonary nodules in positron emission tomography-computed tomography imaging using a two-stage multimodal learning approach." Quantitative Imaging in Medicine and Surgery 14(8): 5526–5540.
Li, X. R., J. J. Jin, Y. Yu, X. H. Wang, Y. Guo and H. Z. Sun (2021). "PET-CT radiomics by integrating primary tumor and peritumoral areas predicts E-cadherin expression and correlates with pelvic lymph node metastasis in early-stage cervical cancer." Eur Radiol 31(8): 5967–5979.
Li, Y., D. Han and C. Shen (2024). "Prediction of the axillary lymph-node metastatic burden of breast cancer by (18)F-FDG PET/CT-based radiomics." BMC Cancer 24(1): 704.
Li, Y., M. R. Imami, L. Zhao, A. Amindarolzarbi, E. Mena, J. Leal, J. Chen, A. Gafita, A. F. Voter, X. Li, Y. Du, C. Zhu, P. L. Choyke, B. Zou, Z. Jiao, S. P. Rowe, M. G. Pomper and H. X. Bai (2024). "An Automated Deep Learning-Based Framework for Uptake Segmentation and Classification on PSMA PET/CT Imaging of Patients with Prostate Cancer." Journal of Imaging Informatics in Medicine 37(5): 2206–2215.
Li, Y., J. Jiang, J. Lu, J. Jiang, H. Zhang and C. Zuo (2019). "Radiomics: a novel feature extraction method for brain neuron degeneration disease using 18F-FDG PET imaging and its implementation for Alzheimer's disease and mild cognitive impairment." Therapeutic advances in neurological disorders 12: 1756286419838682.
Lin, Q., T. Li, C. Cao, Y. Cao, Z. Man and H. Wang (2021). "Deep learning based automated diagnosis of bone metastases with SPECT thoracic bone images." Scientific Reports 11(1): 4223.

Liu, H., Y. Cui, C. Chang, Z. Zhou, Y. Zhang, C. Ma, Y. Yin and R. Wang (2024). "Development and validation of a (18)F-FDG PET/CT radiomics nomogram for predicting progression free survival in locally advanced cervical cancer: a retrospective multi center study." BMC Cancer 24(1): 150.
Liu, J., C. Sui, H. Bian, Y. Li, Z. Wang, J. Fu, L. Qi, K. Chen, W. Xu and X. Li (2024). "Radiomics based on (18)F-FDG PET/CT for prediction of pathological complete response to neoadjuvant therapy in non-small cell lung cancer." Front Oncol 14: 1425837.
Lohmann, P., M. A. Elahmadawy, R. Gutsche, J. M. Werner, E. K. Bauer, G. Ceccon, M. Kocher, C. W. Lerche, M. Rapp, G. R. Fink, N. J. Shah, K. J. Langen and N. Galldiks (2020). "FET PET Radiomics for Differentiating Pseudoprogression from Early Tumor Progression in Glioma Patients Post-Chemoradiation." Cancers (Basel) 12(12).
Lucia, F., T. Louis, F. Cousin, V. Bourbonne, D. Visvikis, C. Mievis, N. Jansen, B. Duysinx, R. Le Pennec, M. Nebbache, M. Rehn, M. Hamya, M. Geier, P.-Y. Salaun, U. Schick, M. Hatt, P. Coucke, R. Hustinx and P. Lovinfosse (2024). "Multicentric development and evaluation of [18F]FDG PET/CT and CT radiomic models to predict regional and/or distant recurrence in early-stage non-small cell lung cancer treated by stereotactic body radiation therapy." European Journal of Nuclear Medicine and Molecular Imaging 51(4): 1097–1108.
Luo, L., X. Wang, H. Xie, H. Liang, J. Gao, Y. Li, Y. Xia, M. Zhao, F. Shi, C. Shen and X. Duan (2024). "Role of [18F]-PSMA-1007 PET radiomics for seminal vesicle invasion prediction in primary prostate cancer." Computers in Biology and Medicine 183: 109249.
Lv, W., S. Ashrafinia, J. Ma, L. Lu and A. Rahmim (2020). "Multi-Level Multi-Modality Fusion Radiomics: Application to PET and CT Imaging for Prognostication of Head and Neck Cancer." IEEE Journal of Biomedical and Health Informatics 24(8): 2268–2277.
Lv, W., H. Feng, D. Du, J. Ma and L. Lu (2021). "Complementary Value of Intra- and Peri-Tumoral PET/CT Radiomics for Outcome Prediction in Head and Neck Cancer." IEEE Access 9: 81818–81827.
Lv, W., Z. Zhou, J. Peng, L. Peng, G. Lin, H. Wu, H. Xu and L. Lu (2023). "Functional-structural sub-region graph convolutional network (FSGCN): Application to the prognosis of head and neck cancer with PET/CT imaging." Computer Methods and Programs in Biomedicine 230: 107341.
Ma, B., J. Guo, A. De Biase, L. V. van Dijk, P. M. A. van Ooijen, J. A. Langendijk, S. Both and N. M. Sijtsema (2024). "PET/CT based transformer model for multi-outcome prediction in oropharyngeal cancer." Radiotherapy and Oncology 197.
Ma, B., Y. Li, H. Chu, W. Tang, L. R. De la O Arévalo, J. Guo, P. van Ooijen, S. Both, J. A. Langendijk, L. V. van Dijk and N. M. Sijtsema (2023). Deep Learning and Radiomics Based PET/CT Image Feature Extraction from Auto Segmented Tumor Volumes for Recurrence-Free Survival Prediction in Oropharyngeal Cancer Patients. Head and Neck Tumor Segmentation and Outcome Prediction, Cham, Springer Nature Switzerland.
Maes, J., S. Gesquière, A. Maes, M. Sathekge and C. Van de Wiele (2024). "Prostate-Specific Membrane Antigen-Positron Emission Tomography-Guided Radiomics and Machine Learning in Prostate Carcinoma." Cancers (Basel) 16(19).
Mapelli, P., C. Bezzi, F. Muffatti, S. Ghezzo, C. Canevari, P. Magnani, M. Schiavo Lena, A. Battistella, P. Scifo, V. Andreasi, S. Partelli, A. Chiti, M. Falconi and M. Picchio (2024). "Preoperative assessment of lymph nodal metastases with [68Ga]Ga-DOTATOC PET radiomics for improved surgical planning in well-differentiated pancreatic neuroendocrine tumours." European Journal of Nuclear Medicine and Molecular Imaging 51(9): 2774–2783.
Meng, M., B. Gu, L. Bi, S. Song, D. D. Feng and J. Kim (2022). "DeepMTS: Deep Multi-Task Learning for Survival Prediction in Patients With Advanced Nasopharyngeal Carcinoma Using Pretreatment PET/CT." IEEE Journal of Biomedical and Health Informatics 26(9): 4497–4507.
Mirshahvalad, S. A., C. Ortega, U. Metser, R. Jang, E. Chen, D. M. Jiang, J. Yeung, R. Wong and P. Veit-Haibach (2024). "<strong>Value of pre-operative [18F]F-FDG PET/CT in gastroesophageal cancer patientsâ€™ metastatic status prediction and overall survival prognostication: A radiomics study</strong>." Journal of Nuclear Medicine 65(supplement 2): 242212–242212.
Molin, K., N. Barry, S. Gill, G. M. Hassan, R. J. Francis, J. S. L. Ong, M. A. Ebert and J. Kendrick (2025). "Evaluating the prognostic value of radiomics and clinical features in metastatic prostate cancer using [68Ga]Ga-PSMA-11 PET/CT." Physical and Engineering Sciences in Medicine 48(1): 329–341.
Mu, W., L. Jiang, Y. Shi, I. Tunali, J. E. Gray, E. Katsoulakis, J. Tian, R. J. Gillies and M. B. Schabath (2021). "Non-invasive measurement of PD-L1 status and prediction of immunotherapy response using deep learning of PET/CT images." Journal for immunotherapy of cancer 9(6).
Mu, W., L. Jiang, J. Zhang, Y. Shi, J. E. Gray, I. Tunali, C. Gao, Y. Sun, J. Tian, X. Zhao, X. Sun, R. J. Gillies and M. B. Schabath (2020). "Non-invasive decision support for NSCLC treatment using PET/CT radiomics." Nat Commun 11(1): 5228.
Munir, M. A., R. A. Shah, M. Ali, A. A. Laghari, A. Almadhor and T. R. Gadekallu (2025). "Enhancing Gene Mutation Prediction With Sparse Regularized Autoencoders in Lung Cancer Radiomics Analysis." IEEE Access 13: 7407–7425.
Nemoto, H., M. Saito, Y. Satoh, T. Komiyama, K. Marino, S. Aoki, H. Suzuki, N. Sano, H. Nonaka, H. Watanabe, S. Funayama and H. Onishi (2024). "Evaluation of the performance of both machine learning models using PET and CT radiomics for predicting recurrence following lung stereotactic body radiation therapy: A single-institutional study." Journal of Applied Clinical Medical Physics 25(7): e14322.
Nensa, F., A. Demircioglu and C. Rischpler (2019). "Artificial Intelligence in Nuclear Medicine." J Nucl Med 60(Suppl 2): 29s–37s.
Nikkuni, Y., H. Nishiyama and T. Hayashi (2024). "Prediction of Histological Grade of Oral Squamous Cell Carcinoma Using Machine Learning Models Applied to 18F-FDG-PET Radiomics." Biomedicines 12(7): 1411.
Nobashi, T., C. Zacharias, J. K. Ellis, V. Ferri, M. E. Koran, B. L. Franc, A. Iagaru and G. A. Davidzon (2020). "Performance Comparison of Individual and Ensemble CNN Models for the Classification of Brain 18F-FDG-PET Scans." Journal of Digital Imaging 33(2): 447–455.

Noortman, W. A., D. Vriens, L. F. de Geus-Oei, C. H. Slump, E. H. Aarntzen, A. van Berkel, H. Timmers and F. H. P. van Velden (2022). "[(18)F]FDG-PET/CT radiomics for the identification of genetic clusters in pheochromocytomas and paragangliomas." Eur Radiol 32(10): 7227–7236.
Öğülmüş, F. E., Y. Almalıoğlu, M. Ö. Tamam, B. Yıldırım, E. Uysal, Ç. Numanoğlu, H. Özçevik, A. F. Tekin and M. Turan (2025). "Integrating PET/CT, radiomics and clinical data: An advanced multi-modal approach for lymph node metastasis prediction in prostate cancer." Computers in Biology and Medicine 184: 109339.
Pan, K., F. Yao, W. Hong, J. Xiao, S. Bian, D. Zhu, Y. Yuan, Y. Zhang, Y. Zhuang and Y. Yang (2024). "Multimodal radiomics based on 18F-Prostate-specific membrane antigen-1007 PET/CT and multiparametric MRI for prostate cancer extracapsular extension prediction." Br J Radiol 97(1154): 408–414.
Pasini, G., A. Stefano, C. Mantarro, S. Richiusa, A. Comelli, G. I. Russo, M. G. Sabini, S. Cosentino, M. Ippolito and G. Russo (2024). "A Robust [18F]-PSMA-1007 Radiomics Ensemble Model for Prostate Cancer Risk Stratification." Journal of Imaging Informatics in Medicine.
Peng, J., L. Peng, Z. Zhou, X. Han, H. Xu, L. Lu and W. Lv (2024). "Multi-Level fusion graph neural network: Application to PET and CT imaging for risk stratification of head and neck cancer." Biomedical Signal Processing and Control 92: 106137.
Pepponi, M., V. Berti, E. Fasciglione, F. Montanini, L. Canu, F. Hubele, E. Abenavoli, V. Briganti, E. Rapizzi, A. Charpiot, D. Taieb, K. Pacak, B. Goichot and A. Imperiale (2024). "[(68)Ga]DOTATOC PET-derived radiomics to predict genetic background of head and neck paragangliomas: a pilot investigation." Eur J Nucl Med Mol Imaging 51(9): 2684–2694.
Piñeiro-Fiel, M., A. Moscoso, V. Pubul, Á. Ruibal, J. Silva-Rodríguez and P. Aguiar (2021). "A Systematic Review of PET Textural Analysis and Radiomics in Cancer." Diagnostics (Basel) 11(2).
Qi, L., X. Li, J. Ni, Y. Du, Q. Gu, B. Liu, J. He and J. Du (2025). "Construction of feature selection and efficacy prediction model for transformation therapy of locally advanced pancreatic cancer based on CT, (18)F-FDG PET/CT, DNA mutation, and CA199." Cancer Cell Int 25(1): 19.
Qian, L. D., L. J. Feng, S. X. Zhang, J. Liu, J. L. Ren, L. Liu, H. Zhang and J. Yang (2023). "(18)F-FDG PET/CT imaging of pediatric peripheral neuroblastic tumor: a combined model to predict the International Neuroblastoma Pathology Classification." Quant Imaging Med Surg 13(1): 94–107.
Qiao, J., B. Liu, J. Xin, S. Shen, H. Ma and S. Pan (2024). "Prediction of Prognosis and Response to Androgen Deprivation Therapy in Intermediate to High-Risk Prostate Cancer Using (18)F-FDG PET/CT Radiomics." Acad Radiol 31(12): 5008–5021.
Rahmim, A., A. Toosi, M. R. Salmanpour, N. Dubljevic, I. Janzen, I. Shiri, R. Yuan, C. Ho, H. Zaidi, C. MacAulay, C. Uribe and F. Yousefirizi (2023). "Tensor radiomics: paradigm for systematic incorporation of multi-flavoured radiomics features." Quant Imaging Med Surg 13(12): 7680–7694.
Rahmim, A. and H. Zaidi (2008). "PET versus SPECT: strengths, limitations and challenges." Nuclear Medicine Communications 29(3).
Ravikumar, K. V., U. Kumaran, B. T. Sree, B. P. K. Reddy, B. Rethaswi and M. G. Lavanya (2023). Design and Implementation of Lung Abnormality Detection Through Sparse Projection Features Enabled Distributed Convolution Network. 2023 Second International Conference on Electrical, Electronics, Information and Communication Technologies (ICEEICT).
Salimi, Y., G. Hajianfar, Z. Mansouri, A. Sanaat, M. Amini, I. Shiri and H. Zaidi (2024). "Organomics: A Concept Reflecting the Importance of PET/CT Healthy Organ Radiomics in Non-Small Cell Lung Cancer Prognosis Prediction Using Machine Learning." Clin Nucl Med 49(10): 899–908.
Salmanpour, M., M. Hosseinzadeh, A. Akbari, K. Borazjani, K. Mojallal, D. Askari, G. Hajianfar, S. M. Rezaeijo, M. M. Ghaemi, A. H. Nabizadeh and A. Rahmim (2022). Prediction of TNM stage in head and neck cancer using hybrid machine learning systems and radiomics features, SPIE.
Salmanpour, M., M. Hosseinzadeh, E. Modiri, A. Akbari, G. Hajianfar, D. Askari, M. Fatan, M. Maghsudi, H. Ghaffari, S. M. Rezaeijo, M. M. Ghaemi and A. Rahmim (2022). Advanced survival prediction in head and neck cancer using hybrid machine learning systems and radiomics features, SPIE.
Salmanpour, M. R., A. Gorji, A. F. Jouzdani, N. Sanati, R. Yuan and A. Rahmim (2024). Exploring Several Novel Strategies to Enhance Prediction of Lung Cancer Survival Time. 2024 IEEE Nuclear Science Symposium (NSS), Medical Imaging Conference (MIC) and Room Temperature Semiconductor Detector Conference (RTSD).
Salmanpour, M. R., M. Hosseinzadeh, S. M. Rezaeijo and A. Rahmim (2023). "Fusion-based tensor radiomics using reproducible features: Application to survival prediction in head and neck cancer." Computer Methods and Programs in Biomedicine 240: 107714.
Salmanpour, M. R., S. M. Rezaeijo, M. Hosseinzadeh and A. Rahmim (2023). "Deep versus Handcrafted Tensor Radiomics Features: Prediction of Survival in Head and Neck Cancer Using Machine Learning and Fusion Techniques." Diagnostics 13(10): 1696.
Shahzadi, I., A. Seidlitz, B. Beuthien-Baumann, A. Zwanenburg, I. Platzek, J. Kotzerke, M. Baumann, M. Krause, E. G. C. Troost and S. Löck (2024). "Radiomics for residual tumour detection and prognosis in newly diagnosed glioblastoma based on postoperative [11C] methionine PET and T1c-w MRI." Scientific Reports 14(1): 4576.
Shen, W.-C., Y.-H. Chou, M.-W. Chung, G.-Z. Wang, J.-M. Lee and Y.-H. Liao (2024). "Feasibility of pretreatment FDG PET radiomics in predicting circumferential margin involvement for esophageal cancer after neoadjuvant concurrent chemoradiotherapy." Journal of Clinical Oncology 42(23_suppl): 203–203.
Siddu, D. M., A. Pawar, G. Lohith and K. Sekar (2022). "Prostate-specific Membrane Antigen Positron Emission Tomography (PSMA-PET) and Gleason grading system based Artificial Intelligence (AI) Model in Diagnosis and Staging of Prostate Cancer." Journal of Precision Oncology 2(2): 113–119.
Siegel, R. L., T. B. Kratzer, A. N. Giaquinto, H. Sung and A. Jemal (2025). "Cancer statistics, 2025." CA Cancer J Clin 75(1): 10–45.

Song, Y., Y. Tian, X. Lu, G. Chen and X. Lv (2024). "Prognostic value of (18)F-FDG PET radiomics and sarcopenia in patients with oral squamous cell carcinoma." Med Phys 51(7): 4907–4921.

Stefano, A., C. Mantarro, S. Richiusa, G. Pasini, M. G. Sabini, S. Cosentino and M. Ippolito (2023). Prediction of High Pathological Grade in Prostate Cancer Patients Undergoing [18F]-PSMA PET/CT: A Preliminary Radiomics Study. Image Analysis and Processing - ICIAP 2023 Workshops: Udine, Italy, September 11–15, 2023, Proceedings, Part II. Udine, Italy, Springer-Verlag: 49–58.

Stüber, A. T., M. M. Heimer, J. Ta, M. P. Fabritius, B. F. Hoppe, G. Sheikh, M. Brendel, L. Unterrainer, P. Jurmeister, A. Tufman, J. Ricke, C. C. Cyran and M. Ingrisch (2025). "Replication study of PD-L1 status prediction in NSCLC using PET/CT radiomics." European Journal of Radiology 183: 111825.

Sui, C., Q. Su, K. Chen, R. Tan, Z. Wang, Z. Liu, W. Xu and X. Li (2024). "(18)F-FDG PET/CT-based habitat radiomics combining stacking ensemble learning for predicting prognosis in hepatocellular carcinoma: a multi-center study." BMC Cancer 24(1): 1457.

Tong, H., J. Sun, J. Fang, M. Zhang, H. Liu, R. Xia, W. Zhou, K. Liu and X. Chen (2022). "A Machine Learning Model Based on PET/CT Radiomics and Clinical Characteristics Predicts Tumor Immune Profiles in Non-Small Cell Lung Cancer: A Retrospective Multicohort Study." Front Immunol 13: 859323.

Trabelsi, M., H. Romdhane, L. Ben Salem and D. Ben-Sellem (2024). "Advanced artificial intelligence framework for T classification of TNM lung cancer in(18)FDG-PET/CT imaging." Biomed Phys Eng Express 10(6).

Trotter, J., A. R. Pantel, B. K. Teo, F. E. Escorcia, T. Li, D. A. Pryma and N. K. Taunk (2023). "Positron Emission Tomography (PET)/Computed Tomography (CT) Imaging in Radiation Therapy Treatment Planning: A Review of PET Imaging Tracers and Methods to Incorporate PET/CT." Adv Radiat Oncol 8(5): 101212.

Tsujimoto, M., A. Teramoto, M. Dosho, S. Tanahashi, A. Fukushima, S. Ota, Y. Inui, R. Matsukiyo, Y. Obama and H. Toyama (2021). "Automated classification of increased uptake regions in bone single-photon emission computed tomography/computed tomography images using three-dimensional deep convolutional neural network." Nuclear Medicine Communications 42(8).

Usha, M. P., G. Kannan and M. Ramamoorthy (2024). "Multimodal Brain Tumor Classification Using Convolutional Tumnet Architecture." Behavioural Neurology 2024(1): 4678554.

Vagenas, T. P., T. L. Economopoulos, C. Sachpekidis, A. Dimitrakopoulou-Strauss, L. Pan, A. Provata and G. K. Matsopoulos (2023). "A Decision Support System for the Identification of Metastases of Metastatic Melanoma Using Whole-Body FDG PET/CT Images." IEEE Journal of Biomedical and Health Informatics 27(3): 1397–1408.

Vansteenkiste, J. F. and S. S. Stroobants (2006). "PET Scan in Lung Cancer: Current Recommendations and Innovation." Journal of Thoracic Oncology 1(1): 71–73.

Vedaei, F., N. Mashhadi, M. Alizadeh, G. Zabrecky, D. Monti, N. Wintering, E. Navarreto, C. Hriso, A. B. Newberg and F. B. Mohamed (2024). "Deep learning-based multimodality classification of chronic mild traumatic brain injury using resting-state functional MRI and PET imaging." Frontiers in Neuroscience Volume 17 - 2023.

Vial, A., D. Stirling, M. Field, M. Ros, C. Ritz, M. Carolan, L. Holloway and A. A. Miller (2018). "The role of deep learning and radiomic feature extraction in cancer-specific predictive modelling: a review." Translational Cancer Research 7(3): 803–816.

Wang, B., C. Bao, X. Wang, Z. Wang, Y. Zhang, Y. Liu, R. Wang and X. Han (2024). "Inter-equipment validation of PET-based radiomics for predicting EGFR mutation statuses in patients with non-small cell lung cancer." Clin Radiol 79(8): 571–578.

Wang, B., T. Hu, R. Shen, L. Liu, J. Qiao, R. Zhang and Z. Zhang (2025). "A 18F-FDG PET/CT based radiomics nomogram for predicting disease-free survival in stage II/III colorectal adenocarcinoma." Abdominal Radiology 50(1): 64–77.

Wang, H., Y. Chen, J. Qiu, J. Xie, W. Lu, J. Ma and M. Jia (2024). "Machine learning based on SPECT/CT to differentiate bone metastasis and benign bone lesions in lung malignancy patients." Med Phys 51(4): 2578–2588.

Wang, M., M. Peng, X. Yang, Y. Zhang, T. Wu, Z. Wang and K. Wang (2024). "Preoperative prediction of microsatellite instability status: development and validation of a pan-cancer PET/CT-based radiomics model." Nuclear Medicine Communications 45(5).

Wang, N., M. Dai, F. Jing, Y. Liu, Y. Zhao, Z. Zhang, J. Wang, J. Zhang, Y. Wang and X. Zhao (2025). "Value of (18)F-FDG PET/CT-based radiomics features for differentiating primary lung cancer and solitary lung metastasis in patients with colorectal adenocarcinoma." Int J Radiat Biol 101(1): 56–64.

Wang, N., S. Qu, W. Kong, Q. Hua, Z. Hong, Z. Liu and Y. Shi (2024). "Establishment and validation of novel predictive models to predict bone metastasis in newly diagnosed prostate adenocarcinoma based on single-photon emission computed tomography radiomics." Ann Nucl Med 38(9): 734–743.

Wang, X. and Z. Lu (2021). "Radiomics Analysis of PET and CT Components of (18)F-FDG PET/CT Imaging for Prediction of Progression-Free Survival in Advanced High-Grade Serous Ovarian Cancer." Front Oncol 11: 638124.

Wang, X., C. Xu, M. Grzegorzek and H. Sun (2022). "Habitat radiomics analysis of pet/ct imaging in high-grade serous ovarian cancer: Application to Ki-67 status and progression-free survival." Front Physiol 13: 948767.

Wang, Y., E. Lombardo, M. Avanzo, S. Zschaek, J. Weingärtner, A. Holzgreve, N. L. Albert, S. Marschner, G. Fanetti, G. Franchin, J. Stancanello, F. Walter, S. Corradini, M. Niyazi, J. Lang, C. Belka, M. Riboldi, C. Kurz and G. Landry (2022). "Deep learning based time-to-event analysis with PET, CT and joint PET/CT for head and neck cancer prognosis." Computer Methods and Programs in Biomedicine 222: 106948.

Wang, Y., G. Yang, X. Gao, L. Li, H. Zhu and H. Yi (2024). "Subregion-specific (18)F-FDG PET-CT radiomics for the pre-treatment prediction of EGFR mutation status in solid lung adenocarcinoma." Am J Nucl Med Mol Imaging 14(2): 134–143.

Wang, Y., H. Zhao, P. Fu, L. Tian, Y. Su, Z. Lyu, W. Gu, Y. Wang, S. Liu, X. Wang, H. Zheng, J. Du and R. Zhang (2024). "Preoperative prediction of lymph node metastasis in colorectal cancer using (18)F-FDG PET/CT peritumoral radiomics analysis." Med Phys 51(8): 5214–5225.

Wei, W., G. Jia, Z. Wu, T. Wang, H. Wang, K. Wei, C. Cheng, Z. Liu and C. Zuo (2023). "A multidomain fusion model of radiomics and deep learning to discriminate between PDAC and AIP based on (18)F-FDG PET/CT images." Jpn J Radiol 41(4): 417–427.

Wilk, A., D. Borys, K. Fujarewicz, A. d'Amico, R. Suwiński and A. Świerniak (2022). Potential of Radiomics Features for Predicting Time to Metastasis in NSCLC. Intelligent Information and Database Systems, Cham, Springer Nature Switzerland.

Wu, D., Y. Li, M. Zhou, F. Gong and J. Li (2024). "Deep learning-based characterization of pathological subtypes in lung invasive adenocarcinoma utilizing 18F-deoxyglucose positron emission tomography imaging." BMC Cancer 24(1): 1229.

Xing, F., M. Silosky, D. Ghosh and B. B. Chin (2024). "Location-Aware Encoding for Lesion Detection in $^{68}$Ga-DOTATATE Positron Emission Tomography Images." IEEE Transactions on Biomedical Engineering 71(1): 247–257.

Xu, H., N. Abdallah, J.-M. Marion, P. Chauvet, C. Tauber, T. Carlier, L. Lu and M. Hatt (2023). "Radiomics prognostic analysis of PET/CT images in a multicenter head and neck cancer cohort: investigating ComBat strategies, sub-volume characterization, and automatic segmentation." European Journal of Nuclear Medicine and Molecular Imaging 50(6): 1720–1734.

Xu, L., G. Huang, Y. Wang, G. Huang, J. Liu and R. Chen (2024). "2-[(18)F]FDG PET-based quantification of lymph node metabolic heterogeneity for predicting lymph node metastasis in patients with colorectal cancer." Eur J Nucl Med Mol Imaging 51(6): 1729–1740.

Xu, S., C. Zhu, M. Wu, S. Gu, Y. Wu, S. Cheng, C. Wang, Y. Zhang, W. Zhang, W. Shen, J. Yang, X. Yang and Y. Wang (2025). "Artificial intelligence algorithm for preoperative prediction of FIGO stage in ovarian cancer based on clinical features integrated 18F-FDG PET/CT metabolic and radiomics features." Journal of Cancer Research and Clinical Oncology 151(2): 87.

Xue, X. Q., W. J. Yu, X. Shi, X. L. Shao and Y. T. Wang (2022). "(18)F-FDG PET/CT-based radiomics nomogram for the preoperative prediction of lymph node metastasis in gastric cancer." Front Oncol 12: 911168.

Yang, F., C. Wang, J. Shen, Y. Ren, F. Yu, W. Luo and X. Su (2025). "End-to-end [18F]PSMA-1007 PET/CT radiomics-based pipeline for predicting ISUP grade group in prostate cancer." Abdominal Radiology 50(4): 1641–1652.

Yang, L., H. Ding, X. Gao, Y. Xu, S. Xu and K. Wang (2024). "Can we skip invasive biopsy of sentinel lymph nodes? A preliminary investigation to predict sentinel lymph node status using PET/CT-based radiomics." BMC Cancer 24(1): 1316.

Yang, M., X. Li, C. Cai, C. Liu, M. Ma, W. Qu, S. Zhong, E. Zheng, H. Zhu, F. Jin and H. Shi (2024). "[18F]FDG PET-CT radiomics signature to predict pathological complete response to neoadjuvant chemoimmunotherapy in non-small cell lung cancer: a multicenter study." European Radiology 34(7): 4352–4363.

Yang, S., W. Zhang, C. Liu, C. Li and K. Hua (2024). "Predictive value and potential association of PET/CT radiomics on lymph node metastasis of cervical cancer." Ann Med Surg (Lond) 86(2): 805–810.

Yang, Y., B. Zheng, Y. Li, Y. Li and X. Ma (2023). "Computer-aided diagnostic models to classify lymph node metastasis and lymphoma involvement in enlarged cervical lymph nodes using PET/CT." Medical Physics 50(1): 152–162.

Yoon, H., W. H. Choi, M. W. Joo, S. Ha and Y.-A. Chung (2023) "SPECT/CT Radiomics for Differentiating between Enchondroma and Grade I Chondrosarcoma." Tomography 9, 1868–1875 DOI: 10.3390/tomography9050148.

Yu, L., Z. Zhang, H. Yi, J. Wang, J. Li, X. Wang, H. Bai, H. Ge, X. Zheng, J. Ni, H. Qi, Y. Guan, W. Xu, Z. Zhu, L. Xing, A. Dekker, L. Wee, A. Traverso, Z. Ye and Z. Yuan (2024). "A PET/CT radiomics model for predicting distant metastasis in early-stage non-small cell lung cancer patients treated with stereotactic body radiotherapy: a multicentric study." Radiat Oncol 19(1): 10.

Yu, Y., J. Zhu, S. Sang, Y. Yang, B. Zhang and S. Deng (2024). "Application of 18F-FDG PET/CT imaging radiomics in the differential diagnosis of single-nodule pulmonary metastases and second primary lung cancer in patients with colorectal cancer." J Cancer Res Ther 20(2): 599–607.

Yuan, P., Z.-H. Huang, Y.-H. Yang, F.-C. Bao, K. Sun, F.-F. Chao, T.-T. Liu, J.-J. Zhang, J.-M. Xu, X.-N. Li, F. Li, T. Ma, H. Li, Z.-H. Li, S.-F. Zhang, J. Hu and Y. Qi (2024). "A 18F-FDG PET/CT-based deep learning-radiomics-clinical model for prediction of cervical lymph node metastasis in esophageal squamous cell carcinoma." Cancer Imaging 24(1): 153.

Yubo, W., L. Qiang, Z. Shaofang, Z. Xianwu, Z. Bowen, C. Yongchun and M. Zhengxing (2024). "Automated Diagnosis of Bone Metastasis by Classifying Bone Scintigrams Using a Self-defined Deep Learning Model." Current Medical Imaging 20: 1–13.

Zhai, W., X. Li, T. Zhou, Q. Zhou, X. Lin, X. Jiang, Z. Zhang, Q. Jin, S. Liu and L. Fan (2025). "A machine learning-based 18F-FDG PET/CT multi-modality fusion radiomics model to predict Mediastinal-Hilar lymph node metastasis in NSCLC: a multi-centre study." Clinical Radiology 83: 106832.

Zhang, D., B. Zheng, L. Xu, Y. Wu, C. Shen, S. Bao, Z. Tan and C. Sun (2024). "A radiomics-boosted deep-learning for risk assessment of synchronous peritoneal metastasis in colorectal cancer." Insights Imaging 15(1): 150.

Zhang, M.-X., P.-F. Liu, M.-D. Zhang, P.-G. Su, H.-S. Shang, J.-T. Zhu, D.-Y. Wang, X.-Y. Ji and Q.-M. Liao (2025). "Deep learning in nuclear medicine: from imaging to therapy." Annals of Nuclear Medicine.

Zhang, Y., H. Liu, C. Chang, Y. Yin and R. Wang (2024). "Machine learning for differentiating lung squamous cell cancer from adenocarcinoma using Clinical-Metabolic characteristics and 18F-FDG PET/CT radiomics." PLOS ONE 19(4): e0300170.

Zhao, H., Y. Su, Z. Lyu, L. Tian, P. Xu, L. Lin, W. Han and P. Fu (2024). "Non-invasively Discriminating the Pathological Subtypes of Non-small Cell Lung Cancer with Pretreatment 18F-FDG PET/CT Using Deep Learning." Academic Radiology 31(1): 35–45.

Zhao, H., C. Zheng, H. Zhang, M. Rao, Y. Li, D. Fang, J. Huang, W. Zhang and G. Yuan (2023). "Diagnosis of thyroid disease using deep convolutional neural network models applied to thyroid scintigraphy images: a multicenter study." Front Endocrinol (Lausanne) 14: 1224191.

Zhao, S., X. Chen, Z. zheng and Q. Lin (2021). Classifying SPECT Bone Metastasis Images in Grayscale Format with VGGNets. 2021 International Conference on Communications, Information System and Computer Engineering (CISCE).

Zhen, J., J. Chen and Z. Huang (2021). "Radiomics for the prediction of dementia conversion in patients with mild cognitive impairment: a multi-modal study." Aging (Albany NY) 13: 18607–18622.

Zheng, J., Y. Hao, Y. Guo, M. Du, P. Wang and J. Xin (2024). "An 18F-FDG-PET/CT-based radiomics signature for estimating malignance probability of solitary pulmonary nodule." Clin Respir J 18(5): e13751.
Zhi, H., Y. Xiang, C. Chen, W. Zhang, J. Lin, Z. Gao, Q. Shen, J. Shao, X. Yang, Y. Yang, X. Chen, J. Zheng, M. Lu, B. Pan, Q. Dong, X. Shen and C. Ma (2024). "Development and validation of a machine learning-based (18)F-fluorodeoxyglucose PET/CT radiomics signature for predicting gastric cancer survival." Cancer Imaging 24(1): 99.
Zhong, J., R. Frood, P. Brown, H. Nelstrop, R. Prestwich, G. McDermott, S. Currie, S. Vaidyanathan and A. F. Scarsbrook (2021). "Machine learning-based FDG PET-CT radiomics for outcome prediction in larynx and hypopharynx squamous cell carcinoma." Clinical Radiology 76(1): 78.e79–78.e17.
Zhong, S., Y. Wu, Z. Liu, Z. Pan, B. Huang and Q. Yang (2022). Automatic Detection of Prostate Cancer Systemic Lesions Based on Deep Learning and 68Ga-PSMA-11 PET/CT. 2022 IEEE 35th International Symposium on Computer-Based Medical Systems (CBMS).
Zhou, W., J. Wen, Q. Huang, Y. Zeng, Z. Zhou, Y. Zhu, L. Chen, Y. Guan, F. Xie, D. Zhuang and T. Hua (2024). "Development and validation of clinical-radiomics analysis for preoperative prediction of IDH mutation status and WHO grade in diffuse gliomas: a consecutive l-[methyl-11C] methionine cohort study with two PET scanners." European Journal of Nuclear Medicine and Molecular Imaging 51(5): 1423–1435.
Zhou, Y., J. Zhou, X. Cai, S. Ge, S. Sang, Y. Yang, B. Zhang and S. Deng (2024). "Integrating (18)F-FDG PET/CT radiomics and body composition for enhanced prognostic assessment in patients with esophageal cancer." BMC Cancer 24(1): 1402.
Zhu, A., D. Lee and H. Shim (2011). "Metabolic positron emission tomography imaging in cancer detection and therapy response." Semin Oncol 38(1): 55–69.
Zuo, Y., L. Liu, C. Chang, H. Yan, L. Wang, D. Sun, M. Ruan, B. Lei, X. Xia, W. Xie, S. Song and G. Huang (2024). "Value of multi-center (18)F-FDG PET/CT radiomics in predicting EGFR mutation status in lung adenocarcinoma." Med Phys 51(7): 4872–4887.
Zuo, Y., Q. Liu, N. Li, P. Li, Y. Fang, L. Bian, J. Zhang and S. Song (2024). "Explainable (18)F-FDG PET/CT radiomics model for predicting EGFR mutation status in lung adenocarcinoma: a two-center study." J Cancer Res Clin Oncol 150(10): 469.

## Appendix

**Appendix Table A1.** Comprehensive Bias Evaluation Summary. This table reports the number and percentage of studies that received a score of 1 for each evaluation criterion. The criteria assessed include the use of cross-validation, reporting of multiple evaluation metrics, inclusion of diverse datasets, and transparency in model performance reporting. Percentages are rounded to facilitate interpretation.

| Questions (Q) | Number of Studies with Score 1 | Percentage of Studies with Score 1 (%) |
|---|---|---|
| Q1- Was the dataset split (e.g., train, validation, and test sets)? | 221 | 96 |
| Q2- Are the class distributions (e.g., cancer vs. non-cancer regions) imbalanced? | 136 | 59 |
| Q3- Is data augmentation used? | 154 | 67 |
| Q4-Are different MODELS compared systematically? | 221 | 95 |
| Q5- Is cross-validation used properly (e.g., leave-one-center-out)? | 195 | 84 |
| Q6- Have articles referred to hyperparameter tuning methods? | 175 | 76 |
| Q7- Are models evaluated for stability across different initializations or random seeds? | 146 | 63 |
| Q8- Was multicenter data used for training? | 156 | 67 |
| Q9- Was multicenter data used for external testing? | 165 | 71 |
| Q10- At least 6 Evaluation Metrics Reported | 190 | 82 |
| Q11- At least 5 Evaluation Metrics Reported | 196 | 84 |
| Q12- At least 4 Evaluation Metrics Reported | 205 | 88 |
| Q13- At least 3 Evaluation Metrics Reported | 231 | 92 |
| Q14- At least 2Evaluation Metric Reported | 224 | 97 |
| Q15- At least 1 Evaluation Metric Reported? | 231 | 100 |
| Q16- Were outcome definitions (classification, regression, prediction) consistent across all models? | 210 | 91 |
| Q17- Were performance metrics (e.g., for classification: AUC, accuracy, and F1-score; for regression: RMSE and MAE; and for clustering: silhouette score and Davies-Bouldin index) evaluated using the same test sets for all models?" | 224 | 97 |
| Q18- Was there transparency in reporting model performance, including both successes and limitations? | 203 | 88 |
| Q19- Does the dataset include a wide range of cases, not just severe or advanced cases?? | 189 | 82 |
| Q20-Does model performance vary based on socioeconomic factors or geographic locations? | 20 | 9 |
| Q21- Are the eligibility criteria and any treatments received by participants described? | 151 | 65 |
| Q22- Is the handling of missing data described, and its impact on the results discussed? | 68 | 29 |
| Q23- Was multicenter data used for training? | 174 | 75 |
| Q24- Have the harmonization techniques that used for multicenter datasets or images produced by different scanners | 131` | 57 |
| Q25- Were differences in patient populations/races used for training, validation, and testing considered in model comparison? | 45 | 19 |
| Q26- Did the authors discuss the model's generalizability to unseen data, particularly considering that they did not have access to this data themselves and relied on external | 185 | 80 |
| Q27- Did the study provide enough detail to replicate the results (e.g., code, data availability)? | 173 | 75 |
| Q28- Was Ground truth labeling by multiple experts employed | 127 | 55 |
| Q29- Were the scanners in the dataset obtained using different scanners and imaging protocols across various institutions? | 160 | 69 |
| Q30- Was the data source clearly defined and consistent across all models? | 226 | 98 |
| Q31- Was a large sample size used to train, validate, and test each model? (At least # 1000) | 12 | 5 |
| Q32- Were imaging protocols (e.g., acquisition parameters, reconstruction methods) standardized across all studies? | 189 | 82 |
| Q33- Were automated segmentation methods (using DL or thresholding, etc.) for HRF or DRF? | 194 | 83 |
| Q34- Was the impact of image filtering, noise reduction, etc., on model performance discussed properly? | 158 | 68 |
| Q35- Were data preprocessing methods (e.g., normalization, augmentation, resampling) consistently applied across all models? | 208 | 90 |
| Q36- Were appropriate evaluation measurements and metrics taken to avoid data leakage between training and testing phases? | 220 | 95 |
| Q37- Was dimension reduction done independently from the model training process? | 130 | 56 |
| Q38- Were the HRF (e.g., shape, texture) chosen based on their relevance to the clinical outcome? | 194 | 84 |
| Q39- Was objective hyperparameter optimization reported? | 196 | 85 |
| Q40- Were appropriate regularization techniques (e.g., dropout, L2 regularization) used in the DL models to mitigate overfitting? | 158 | 66 |
| Q41- Was there any evidence of overfitting in DL models (e.g., high performance on training vs. test data)? | 85 | 36 |
| Q42- Was the choice of HRF or DRF(Feature selection) justified in the context of the clinical task? | 201 | 87 |
| Q43- Were handcrafted radiomics features extracted from the software/packages standardized by IBSI? | 110 | 48 |
| Q44- Were preprocessing pipelines or software packages fully disclosed for HRF and DL? | 210 | 91 |
| Q45- Was there transparency about the selection of specific hyperparameters or thresholds during model training? | 200 | 87 |
| Q46- DRF vs. HRF | 22 | 10 |
| Q47- DRF vs. DL | 9 | 4 |
| Q48- HRF vs. DL | 20 | 9 |
| Q49- DRF vs. HRF vs. DL | 6 | 3 |
| Q50- Were comparisons between HRF or DRF or DL models conducted using statistical significance tests | 134 | 58 |
| Q51- Were performance differences between models assessed for statistical significance? | 197 | 85 |
| Q52- Was external testing performed on all models using an independent dataset? | 164 | 71 |
| Q53- Were different methods used to handle missing data consistently applied across all models? | 67 | 29 |
| Q54- Did the authors specify or report all model parameters, programming languages, statistical methods, resources (GPU or Azure, …) for ensuring reproducibility? | 203 | 88 |
| Q55- Were cross-validation procedures (e.g., k-fold, leave-one-out) consistently applied to HRF, DRF, and DL models? | 195 | 84 |
| Q56- Were the data stratification used for training and testing, or even N-fold cross-validation? | 203 | 88 |
| Q57- For DRF, whether pre-trained models are used to extract deep radiomics features? | 60 | 26 |
| Q58- For DRF, whether a network trained based on employed data from scratch and then extracts deep features extract this trained model? | 78 | 34 |
| Q59- Was transfer learning used in the DL models, and if so, was it appropriately validated? | 63 | 27 |